\documentclass[aps, pra, 10pt, superscriptaddress, twocolumn, longbibliography, floatfix]{revtex4-1}

\usepackage[nointlimits]{amsmath}
\usepackage{amsthm}
\usepackage{graphicx,nicefrac}
\usepackage{subfigure}
\usepackage{color}
\usepackage{txfonts}
\usepackage{mathtools}
\usepackage{hyperref}
\usepackage{tabularx}
\hypersetup{colorlinks=true, citecolor=blue, linkcolor=blue}
 
\usepackage{physics}

\usepackage{amsfonts,amsmath,amssymb}
\usepackage{array,bm,color}
\usepackage{epsfig,graphicx,nomencl,revsymb4-1,upgreek,url}
\usepackage{hyperref}
\usepackage{algpseudocode}
\usepackage{xspace}
\usepackage{tabularx}
\usepackage[normalem]{ulem}
\usepackage{changes}
\usepackage{enumitem}
\usepackage{booktabs}
\usepackage{mathtools}

\makeatletter
\DeclareFontFamily{OMX}{MnSymbolE}{}
\DeclareSymbolFont{MnLargeSymbols}{OMX}{MnSymbolE}{m}{n}
\SetSymbolFont{MnLargeSymbols}{bold}{OMX}{MnSymbolE}{b}{n}
\DeclareFontShape{OMX}{MnSymbolE}{m}{n}{
    <-6>  MnSymbolE5
   <6-7>  MnSymbolE6
   <7-8>  MnSymbolE7
   <8-9>  MnSymbolE8
   <9-10> MnSymbolE9
  <10-12> MnSymbolE10
  <12->   MnSymbolE12
}{}
\DeclareFontShape{OMX}{MnSymbolE}{b}{n}{
    <-6>  MnSymbolE-Bold5
   <6-7>  MnSymbolE-Bold6
   <7-8>  MnSymbolE-Bold7
   <8-9>  MnSymbolE-Bold8
   <9-10> MnSymbolE-Bold9
  <10-12> MnSymbolE-Bold10
  <12->   MnSymbolE-Bold12
}{}

\let\llangle\@undefined
\let\rrangle\@undefined
\DeclareMathDelimiter{\llangle}{\mathopen}%
                     {MnLargeSymbols}{'164}{MnLargeSymbols}{'164}
\DeclareMathDelimiter{\rrangle}{\mathclose}%
                     {MnLargeSymbols}{'171}{MnLargeSymbols}{'171}
\makeatother

\usepackage{tikz}
\usetikzlibrary{arrows}

\def\bra#1{{\left\langle #1 \right|}}
\def\ket#1{{\left| #1 \right\rangle}}

\newcommand{\superket}[1]{|#1\rrangle}
\newcommand{\superbra}[1]{\llangle #1|}
\newcommand{\superketbra}[2]{\ket{#1}\!\rangle\!\langle\!\bra{#2}}

\DeclareMathOperator*{\E}{\mathbb{E}}

\newcommand{\clbit}[1]{\mathbf{e_{#1}}} 

\newcommand{\pauli}{\mathbf{P}}

\newcommand{\Q}[1]{\texttt{Q#1}}

\newcommand{\layer}[1]{\textsc{#1}}

\newcommand{\supp}{\textrm{supp}}

\newcommand{\mcm}{\textrm{mcm}}

\tikzset{
    vertex/.style={circle,draw,minimum size=1.5em},
    edge/.style={->,> = latex'}
}

\begin{document}
\title{Pauli Noise Learning for Mid-Circuit Measurements}
\author{Jordan Hines}
\thanks{jordanh@berkeley.edu}
\affiliation{Department of 
Physics, University of California, Berkeley, CA 94720}
\affiliation{Quantum Performance Laboratory, Sandia National Laboratories, Livermore, CA 94550}
\author{Timothy Proctor}
\thanks{tjproct@sandia.gov}
\affiliation{Quantum Performance Laboratory, Sandia National Laboratories, Livermore, CA 94550}\begin{abstract} 
      Current benchmarks for mid-circuit measurements (MCMs) are limited in scalability or the types of error they can quantify, necessitating new techniques for quantifying MCM performance. Here, we introduce a theory for learning stochastic Pauli noise in MCMs and use it to create MCM cycle benchmarking, a scalable method for benchmarking MCMs. MCM cycle benchmarking extracts detailed information about the rates of errors in randomly compiled layers of MCMs and Clifford gates, and we demonstrate how its results can be used to quantify correlated errors during MCMs on current quantum hardware. Our method can be integrated with existing Pauli noise learning techniques to scalably characterize the errors in wide classes of circuits containing MCMs.
\end{abstract}
\maketitle
Mid-circuit measurements (MCMs) are a critical component of quantum error correction \cite{Campbell2017-tw, Google_Quantum_AI2023-yr, Bluvstein2023-dp, Krinner2022-tp, Gupta2024-pr} and some quantum algorithms \cite{lu2022measurement, griffiths1996semiclassical, bäumer2023efficient}, but they are a large source of error in contemporary quantum processors \cite{Google_Quantum_AI2023-yr, Bluvstein2023-dp, Krinner2022-tp, Gupta2024-pr, rudinger2022characterizing, strickler2022characterizing, govia2023randomized, gaebler2021suppression}. There are currently few techniques capable of quantifying the errors in MCMs, and those that can are limited in scope or scalability. Full tomography \cite{rudinger2022characterizing, strickler2022characterizing} is only feasible for small systems, whereas more scalable methods based on randomized benchmarking \cite{govia2023randomized, gaebler2021suppression} only assess MCM-induced crosstalk errors. 

Pauli noise learning   \cite{flammia2020efficient, harper2020efficient, harper2021fast, chen_learnability_2023, flammia2022averaged} is a family of methods for partial characterization of gates offering an intermediate approach between full tomography and randomized benchmarking. These techniques are widely used, enable error mitigation \cite{temme2017error, vandenBerg2023, gupta_probabilistic_2023}, and can characterize some errors in syndrome extraction circuits \cite{harper2023learning, hockings2024scalable}. However, because noise in MCMs cannot be tailored into the same stochastic Pauli error structure as gate error \cite{wallman2016noise, beale2023randomized}, existing Pauli noise learning techniques cannot fully characterize MCMs \cite{hashim2024quasiprobabilistic, gupta_probabilistic_2023}.

In this letter, we introduce and demonstrate Pauli noise learning techniques for MCMs. We introduce theory for Pauli noise learning of uniform stochastic instruments (USIs) \cite{mclaren2023stochastic}, a noise model for circuit layers containing MCMs that can be enforced using a recent extension of randomized compilation \cite{wallman2016noise, beale2023randomized}. We apply this theory to create MCM cycle benchmarking (MCM-CB), a scalable protocol for estimating the fidelity of MCM-containing layers that generalizes cycle benchmarking \cite{erhard2019characterizing, carignan-dugas_estimating_2023}. We then show how jointly characterizing multi-qubit gates and MCMs with MCM-CB enables learning more error parameters than individual component characterization.

\begin{figure}[h!]
    \centering\includegraphics{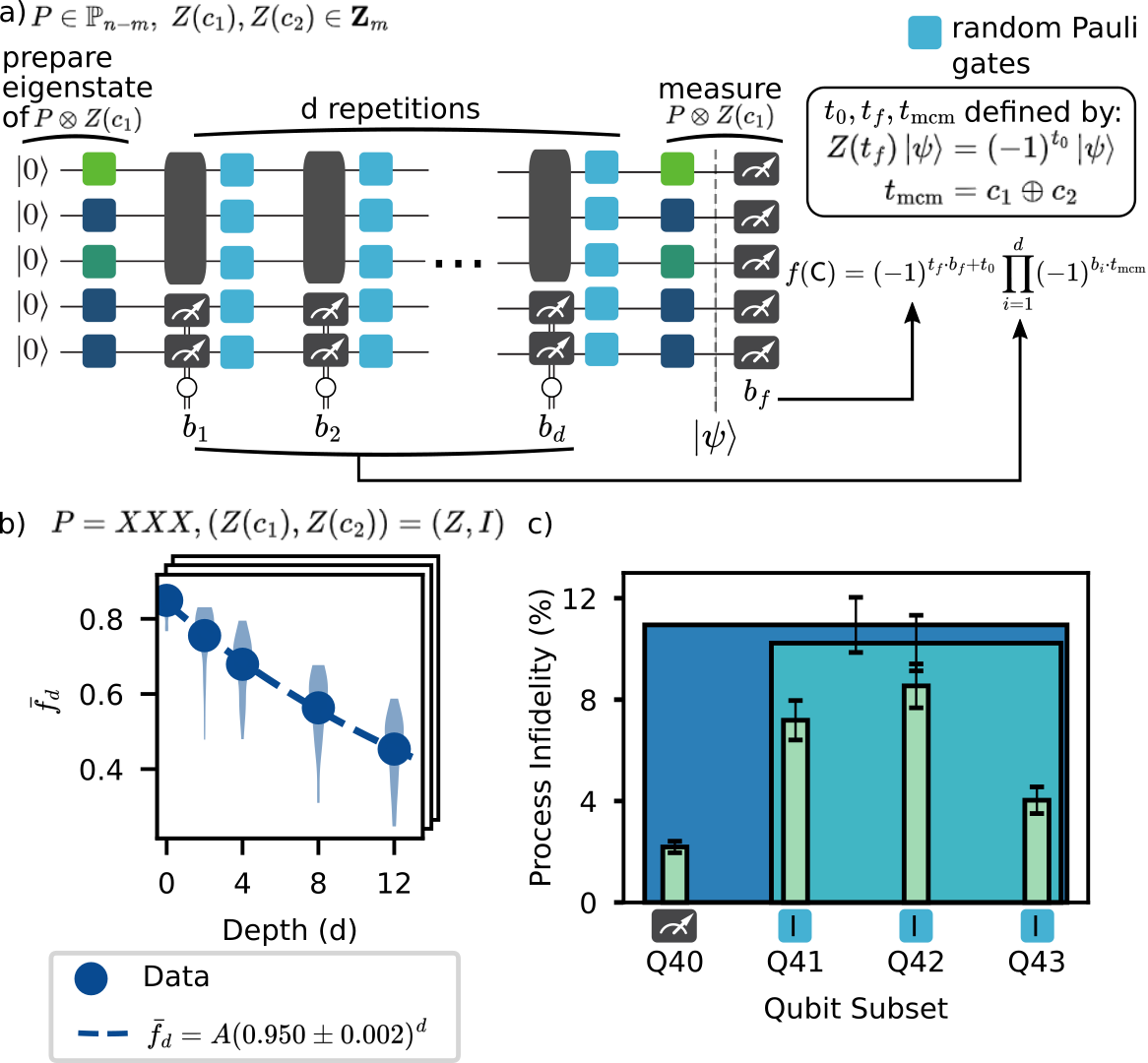}
    \caption{\textbf{Estimating MCM layer fidelity with MCM-CB.} Our protocol runs random subexperiments parameterized by Pauli operators $P$, $Z(c_1)$, and $Z(c_2)$. (a) Each subexperiment runs circuits with $d$ repetitions of a randomly compiled layer $\overline{\layer{L}}$ and measures $P \otimes Z(c_1)$. This measurement result is modified by the MCM results and $Z^{\otimes (c_1 \oplus c_2)}$ to compute $f(C)$, which is averaged over circuits of depth $d$ to obtain $\overline{f}_d$. 
    (b) Each $\overline{f}_d$ is fit to an exponential, and fit decay constants from many subexperiments are used to estimate the fidelity of $\overline{\layer{L}}$. (c) Infidelity estimates from MCM-CB on \texttt{ibm\_osaka} for a layer consisting of one MCM and 3 idling qubits. Error on the idling qubits is dominant, and the sum of the individual idling qubit infidelities is much greater than the process infidelity of the 3-qubit subsystem, indicative of correlated errors.}
    \label{fig:fig1}
\end{figure}    

\vspace{0.2cm}

\noindent
\textbf{Mid-circuit measurements---}We use $\pauli_{n}$ to denote the $n$-qubit Pauli operators (without signs).
For $a \in \mathbb{Z}_2^k$, we use $P(a)$ to denote the Pauli operator $P^{a_1}\otimes P^{a_2} \otimes \dots \otimes P^{a_k}$, called a $P$-type Pauli operator. We denote the set of $m$-qubit $Z$-type Pauli operators by $\mathbf{Z}_m$. We use $\mathcal{P}$ to denote the superoperator representation of $P$. 

An $n$-qubit \emph{MCM layer} $\layer{L}$ is an instruction to perform measurements on $m$ qubits and unitary gates on the remaining $n-m$ qubits. We only consider MCM layers containing computational basis measurements and Clifford gates. Implementing $\layer{L}$ ideally performs the process
\begin{equation}
    \mathcal{L} = \sum_{k  \in \mathbb{Z}_2^m} \mathcal{V} \otimes \superketbra{k}{k} \otimes \clbit{k}, \label{eq:ideal_qi}
\end{equation}
where $\clbit{k}$ denotes the classical state corresponding to MCM outcome $k \in \mathbb{Z}_2^m$, $\superket{k} \equiv \superket{\ket{k}\bra{k}}$, $\superket{\rho}$ is the vector representing operator $\rho$ in Hilbert-Schmidt space [$\mathcal{B}(\mathcal{H})$], and $\mathcal{V}$ is the superoperator representation of the unitary acting on the unmeasured qubits. We will work in the Pauli basis throughout.

An imperfect implementation of $\layer{L}$ can be represented as a \emph{quantum instrument} $\mathcal{M}: \mathcal{B}(\mathcal{H}) \rightarrow \mathcal{B}(\mathcal{H}) \otimes \mathbb{Z}_2^m$, where
\begin{equation}
    \mathcal{M} = \sum_{k  \in \mathbb{Z}_2^m} \mathcal{M}_k \otimes \clbit{k}.
\end{equation}
Each $\mathcal{M}_k$ is completely positive and trace non-increasing, and $\sum_{k  \in \mathbb{Z}_2^m} \mathcal{M}_k$ is trace preserving \cite{rudinger2022characterizing, strickler2022characterizing}.
The process (i.e., entanglement) fidelity of $\mathcal{M}$ to $\mathcal{L}$ is \cite{mclaren2023stochastic}
\begin{equation}
F(\mathcal{M}) \equiv F(\mathcal{M}, \mathcal{L})  = \left(\sum_{k  \in \mathbb{Z}_2^m} \sqrt{F(\mathcal{M}_k,\mathcal{V} \otimes \superketbra{k}{k})}\right)^2.\label{eq:qi_fidelity}   
\end{equation}

\vspace{0.2cm}

\noindent
\textbf{USIs---}Our method learns the error of USIs, a class of quantum instruments with only stochastic error. USIs have the form
\begin{equation}
    \overline{\mathcal{M}} 
     = \sum_{k \in \mathbb{Z}_2^m} \overline{\mathcal{M}}_k \otimes \clbit{k}, \label{eq:usi}
\end{equation}
where
\begin{equation}    \overline{\mathcal{M}}_k =  \sum_{a,b 
    \in \mathbb{Z}_2^m} \mathcal{T}_{a,b}\mathcal{V} \otimes \mathcal{X}(b) \superketbra{k}{k}\mathcal{X}(a)
\end{equation}
is the sum of $4^{m}$ terms labelled by each possible pre- and post-measurement bit flip error pattern (encoded by $a,b \in \mathbb{Z}_2^m$) on the measured qubits. The $\mathcal{T}_{a,b}$ are unnormalized $(n-m)$-qubit stochastic Pauli channels acting on the unmeasured qubits satisfying
$\sum_{a,b \in \mathbb{Z}_2^m} \textrm{Tr}(\mathcal{T}_{a,b})=1$ \cite{beale2023randomized}. $\overline{\mathcal{M}}$'s process fidelity is $F(\overline{\mathcal{M}}) = \Tr(\mathcal{T}_{0,0})$, which is the probability of no error \cite{mclaren2023stochastic}.

Our method uses \emph{randomized compiling} for MCM layers \cite{wallman2016noise, beale2023randomized} to enforce a USI model. A randomly compiled MCM layer $\layer{L}$ is a composite layer $\overline{\layer{L}} = \layer{T}_0\layer{L}\layer{T}_0'$ (read left to right), where $\layer{T}_0$ contains uniformly random Pauli gates on each qubit and $\layer{T}_0'$ is $\layer{T}_0$'s inverse composed with a uniformly random layer of $Z$ gates. Each measured qubit in $\layer{L}$ for which $\layer{T}_0$ contains an $X$ or $Y$ gate has its classical MCM outcome is flipped. With this post-processing, $\overline{\layer{L}}$ is equivalent to $\layer{L}$ up to Pauli operations. 
In a multi-layer circuit, adjacent layers of Pauli gates are compiled together. When single-qubit gate error is much smaller than the error in $\mathcal{M}$, assumed hereafter, an imperfect implementation of $\overline{\layer{L}}$ is approximately a USI.

\vspace{0.2cm}

\noindent
\textbf{Theory of learning USIs---}We now introduce our theory for learning USIs. We assume that the unmeasured subsystem's error-free evolution is $\mathcal{V}=\mathbb{I}_{n-m}$ (see the SM for the general case \cite{supp}). To determine a USI, we must learn all of its stochastic Pauli channels $\mathcal{T}_{a,b}$. Stochastic Pauli channels are diagonal in the Pauli transfer matrix (PTM) representation, and existing methods for learning a Pauli channel \cite{flammia2020efficient, harper2020efficient, erhard2019characterizing, harper2021fast} directly estimate each of its PTM's diagonal elements. 
Our technique is also built around estimating individual PTM elements.

The elements of $\overline{\mathcal{M}}_k$'s PTM
are $[\overline{\mathcal{M}}_k]_{P'\otimes Q', P\otimes Q} = \superbra{P' \otimes Q'}\overline{\mathcal{M}}_k\superket{P \otimes Q}$, where we partition the $n$-qubit Pauli operators into Pauli operators $P,P' \in \pauli_{n-m}$ on the unmeasured qubits and $Q, Q' \in \pauli_{m}$ on the measured qubits. We have that
\begin{align}    [\overline{\mathcal{M}}_k]_{P'\otimes Q', P\otimes Q} &= \superbra{P' \otimes Q'}\left(\sum_{a,b \in \mathbb{Z}_2^m}\mathcal{T}_{a,b} \otimes \mathcal{X}(b)\superketbra{k}{k}\mathcal{X}(a)\right)\superket{P \otimes Q} \nonumber \\ &= \sum_{a,b \in \mathbb{Z}_2^m}\superbra{P'}  \mathcal{T}_{a,b} \superket{P}\superbra{Q'}\mathcal{X}(b)\superketbra{k}{k}\mathcal{X}(a)\superket{Q}. \label{eq:PTM-1} 
\end{align}
Because $\superket{k}$ is a computational basis state, $\superbra{k}\mathcal{X}(a)\superket{Q}=0$ for all $Q \in \pauli_m \setminus \mathbf{Z}_m$ and $a \in \mathbb{Z}_2^m$. Furthermore, because each $\mathcal{T}_{a,b}$ is a stochastic Pauli channel, $\superbra{P'} \mathcal{T}_{a,b} \superket{P} = 0$ for all $P' \neq P$. Substituting these equalities into Eq.~\eqref{eq:PTM-1}, we obtain $[\overline{\mathcal{M}}_k]_{P'\otimes Q', P \otimes Q} = 0$ when $P \neq P'$ and/or $Q \notin \mathbf{Z}_m$, and otherwise, letting $Q=Z(c)$, $Q'=Z(c')$, and $\lambda_{a,b,P} = \superbra{P} \mathcal{T}_{a,b} \superket{P}$,
\begin{align}
[\overline{\mathcal{M}}_k]_{P\otimes Z(c'), P \otimes Z(c)}  & = \sum_{a,b \in \mathbb{Z}_2^m} \lambda_{a,b,P}(-1)^{(k\oplus a) \cdot c' + (k \oplus b) \cdot c} \\
     & = (-1)^{k\cdot (c\oplus c')}\tilde{\lambda}_{P,(Z(c), Z(c'))}.\label{eq:ptm_elt}
\end{align}
Here, $b_1 \oplus b_2$ is the bitwise XOR of $b_1, b_2 \in \mathbb{Z}^m_2$, and \footnote{Eq.\eqref{eq:lambda_tilde} is a binary Fourier transform, and therefore $\tilde{\lambda}_{P,(Z(c), Z(c'))}$ can also be thought of as a Fourier coefficient \cite{zhang2024generalized}.}
\begin{equation}
    \tilde{\lambda}_{P,(Z(c), Z(c'))} = \sum_{a, b  \in \mathbb{Z}_2^m} \lambda_{a,b, P} (-1)^{a \cdot c + b \cdot c'}. \label{eq:lambda_tilde}
\end{equation}
Using Eq.~\eqref{eq:ptm_elt}, $\overline{\mathcal{M}}_k$ is 
\begin{align}
    \overline{\mathcal{M}}_k& = \sum_{P \in \pauli_{n-m}}\sum_{c, c' \in \mathbb{Z}_2^{m}}(-1)^{k\cdot (c\oplus c')}\tilde{\lambda}_{P,(Z(c), Z(c'))}\superketbra{P \otimes Z(c')}{P \otimes Z(c)}. \label{eq:m_k}
\end{align}
Therefore, the $(P\otimes Z(c), P \otimes Z(c'))$ element of $\overline{\mathcal{M}}_k$ has value $\pm\tilde{\lambda}_{P,(Z(c), Z(c'))}$ for all $k$, where the sign depends on the MCM result $k$. 

Learning all $\tilde{\lambda}_{P,(Z(c), Z(c'))}$ is sufficient to learn the USI. However, this is complicated because which $\overline{\mathcal{M}}_k$ occurs when $\overline{\mathcal{M}}$ is performed depends on the MCM outcome $k$, which is uniformly random. If we apply $\overline{\mathcal{M}}$ to $\superket{P \otimes Z(c)}$ and ignore the measurement results (tracing over the classical register) we can measure
\begin{align}
     \sum_{k \in \mathbb{Z}_2^m}\superbra{P \otimes Z(c')} \otimes \clbit{k}^{\textrm{T}} \overline{\mathcal{M}}\superket{P \otimes Z(c)} & = \sum_{k \in \mathbb{Z}_2^m}(-1)^{k\cdot (c\oplus c')}\tilde{\lambda}_{P,(Z(c),Z(c'))} \nonumber \\
     & = \delta_{c, c'}\tilde{\lambda}_{P,(Z(c), Z(c'))},
\end{align}
enabling learning the diagonal PTM elements. But to fully learn $\overline{\mathcal{M}}$, we must incorporate the MCM results.

An MCM result ($k$) tells us which $\overline{\mathcal{M}}_k$ occured, enabling computing the $(-1)^{k\cdot (c\oplus c')}$ factors in the applied $\overline{\mathcal{M}}_k$. We can then simulate measuring $(-1)^{k\cdot (c\oplus c')} P \otimes Z(c')$ by measuring $P \otimes Z(c')$ and applying the sign in post-processing. This enables measuring
\begin{align}
    \sum_{k \in \mathbb{Z}_2^m} \superbra{P \otimes Z(c')}   \otimes \clbit{k}^{\textrm{T}} (-1)^{k\cdot (c\oplus c')}\overline{\mathcal{M}}_k\superket{P \otimes Z(c)} & = \tilde{\lambda}_{P,(Z(c), Z(c'))} .
\end{align}

We have shown that $\overline{\mathcal{M}}$ can be parameterized by $\{\tilde{\lambda}_{P,(Z(c), Z(c'))}\}$, and that each each $\tilde{\lambda}_{P,(Z(c), Z(c'))}$ is either a diagonal element of each $\overline{\mathcal{M}}_k$ that can be extracted without using MCM results, or an off-diagonal element of each $\overline{\mathcal{M}}_k$ that can be extracted by applying a $k$-dependent sign in post-processing. We have therefore shown how to reconstruct $\overline{\mathcal{M}}$. However, we may instead want to directly estimate $F(\overline{\mathcal{M}})$ or the rates of individual Pauli errors. We do so by estimating the eigenvalues $\lambda_{a,b,P}$ of $\mathcal{T}_{a,b}$. 
For fixed $P$, the $\tilde{\lambda}_{P,(Z(c), Z(c'))}$ are orthogonal linear combinations of $\{\lambda_{a,b,P}\}_{a,b \in \mathbb{Z}_2^m}$ (see the SM \cite{supp}) and can be used to determine individual $\lambda_{a,b,P}$. In particular, the fidelity of $\overline{\mathcal{M}}$ is $\textrm{Tr}(\mathcal{T}_{0,0})$, and $\mathcal{T}_{0,0}$'s eigenvalues are $\lambda_{0,0,P} = \frac{1}{4^m}\sum_{c,c' \in \mathbb{Z}_2^{m}}\tilde{\lambda}_{P,(Z(c), Z(c'))}$.
Therefore,
\begin{align}
    F(\overline{\mathcal{M}}) = \textrm{Tr}(\mathcal{T}_{0,0}) & = \sum_{P \in \pauli_{n-m}}\sum_{\substack{c,c' \in \mathbb{Z}_2^{m}}}\tilde{\lambda}_{P,(Z(c), Z(c'))}. \label{eq:fidelity_result}
\end{align}

\vspace{0.2cm}

\noindent
\textbf{MCM cycle benchmarking---}We now introduce MCM-CB. MCM-CB estimates the process fidelity of the USI of a randomly compiled MCM layer by running random Pauli noise learning subexperiments that each learn a single parameter of $\overline{\mathcal{M}}$ [Fig.~\ref{fig:fig1}(a)-(b)]. An MCM-CB subexperiment is specified by an $(n-m)$-qubit input Pauli operator $P$, a pair of $m$-qubit $Z$-type Pauli operators $(Z(c_1), Z(c_2))$, a set of depths $d$, and a number of circuits per depth $N$. We require that $d$ is even, $\mathcal{V}^d = \mathbb{I}$, and $\layer{L}$ contains only Clifford gates on the $(n-m)$ unmeasured qubits (techniques for CB of non-Clifford gates \cite{Kim2022, hashim2022optimized} can extend MCM-CB to general gates)

An MCM-CB subexperiment is the following procedure:
\begin{enumerate}
    \item For each depth $d$, generate $N$ circuits $\layer{C} = \layer{L}_0\overline{\layer{L}}^d\layer{L}_{f}$ (read left to right), where
\begin{enumerate}
    \item $\layer{L}_0$ contains single-qubit gates preparing a random tensor product eigenstate of $P \otimes Z(1\cdots 1)$,
    \item $\overline{\layer{L}}$ is the randomized compilation of $\layer{L}$, (i.e., as described above,
    it is preceded and proceeded by random single-qubit Pauli gates, and MCM results are processed to account for these gates)  and
    \item  $\layer{L}_{f}$ contains single-qubit gates that transform $P$ into $Z(t_f)$, where $t_f$ is an ($n-m$)-bit string encoding the support of $P$. $\layer{L}_f$ can be constructed as follows. Let $P = \bigotimes_{i=1}^{n-m} P_i$, where $P_i$ denotes the single-qubit Pauli operator acting on qubit $i$. On qubit $i$, apply $H$ if $P_i=X$, apply $HS^{\dag}$ if $P_i=Y$, and apply $I$ if $P_i=I$ or $Z$.
\end{enumerate} 
When $\layer{C}$ is implemented without errors, it produces a state $\ket{\psi}$ satisfying $\bra{\psi}Z(t_f)\ket{\psi}=(-1)^{t_0}$, where $t_0 \in \mathbb{Z}_2$ depends on $\layer{L}_0$ and the random Pauli operators inserted for randomized compilation.
    \item Run each circuit $\layer{C}$ and compute
\begin{equation}
    f(\layer{C}) = (-1)^{t_f \cdot b_f+t_0}\prod_{i=1}^d (-1)^{b_i \cdot t_{\mcm}}, \label{eq:data_analysis}
\end{equation}
where $t_{\mcm}$ is defined by $Z(c_1)Z(c_2) = Z(t_{\mcm})$, $b_f$ is the $n$-bit final measurement outcome, and $b_i$ is the $m$-bit MCM outcome from layer $i$. Eq.~\eqref{eq:data_analysis} assumes that randomized compiling's classical postprocessing of MCM outcomes has been applied to $b_i$.
\item Compute $\overline{f}_{d} = \frac{1}{N} \sum\limits_{j=1}^N f(\layer{C}_j),$
and fit it to 
\begin{equation}
    \overline{f}_{d} = Ar^d_{P, (Z(c_1),Z(c_2))}. 
\end{equation}
\end{enumerate} 
An MCM-CB subexperiment's result is an estimate $\hat{r}_{P, (Z(c_1),Z(c_2))}$ of $r_{P, (Z(c_1),Z(c_2))}$, where $r_{P, (Z(c_1),Z(c_2))}$ is determined by the PTM elements $\tilde{\lambda}_{P, (Z(c_1), Z(c_2))}$ of $\overline{\mathcal{M}'}$, and  $\overline{\mathcal{M}} = \overline{\mathcal{M}'}(\mathcal{V}\otimes\mathbb{I}_{m})$. Specifically,
\begin{equation}
    r_{P, (Z(c_1),Z(c_2))} = \sqrt[\ell]{\prod_{j=1}^{\ell} \tilde{\lambda}_{\mathcal{V}^j[P], (Z(c_2)^{j-1}Z(c_1), Z(c_2)^{j}Z(c_1))}}, \label{eq:products}
\end{equation} 
where $\ell$ is the smallest positive even integer satisfying $\mathcal{V}^{\ell}=\mathbb{I}_{n-m}$, assuming that $\tilde{\lambda}_{\mathcal{V}^j[P], (Z(c_2)^{j-1}Z(c_1), Z(c_2)^{j}Z(c_1))}> \delta$ for some constant $\delta > 0$ and all $1 \leq j \leq \ell$. One MCM-CB circuit set can be used to estimate $\hat{r}_{P, (Z(c_1),Z(c_2))}$ for all $(Z(c_1), Z(c_2)) \in \mathbf{Z}_m \times \mathbf{Z}_m$, because the circuits are determined by $P$. These $\hat{r}_{P, (Z(c_1),Z(c_2))}$ can be used to estimate $\lambda_{a,b,P}$ by assuming $\tilde{\lambda}_{P, (Z(c_2),Z(c_1))} \approx r_{P, (Z(c_1),Z(c_2))}$.

MCM-CB consists of running many MCM-CB subexperiments to estimate the fidelity of $\overline{\mathcal{M}}$:
\begin{enumerate}
    \item Pick $K$ uniformly random triplets of Pauli operators $(P,Z(c_1), Z(c_2)) \in \pauli_{n-m} \times \mathbf{Z}_m \times \mathbf{Z}_m$.
    \item For each triplet of Pauli operators $(P_{k}, Z(c_{1,k}), Z(c_{2,k}))$, perform an MCM-CB subexperiment with input Pauli $P_k$ and measured subsystem Paulis $(Z(c_{1,k}),Z(c_{2,k}))$, and estimate the decay constant $\hat{r}_{P_{k}, (Z(c_{1, k}), Z(c_{2,k}))}$.
    \item Estimate the process fidelity by
    \begin{equation}
        \hat{F} = \frac{1}{K}\sum_{k=1}^K \hat{r}_{P_{k}, (Z(c_{1, k}), Z(c_{2,k}))}.
    \end{equation}
    \end{enumerate}
When $m=0$, this protocol reduces to CB of Clifford gates. The number of samples $K$ required to estimate $F$ to within a fixed multiplicative precision is independent of $n$, so our method is sample efficient in $n$ (see SM \cite{supp}).

\vspace{0.2cm}

\noindent
\textbf{The fidelity of randomly compiled MCMs---}In the infinite sampling limit, MCM-CB measures a quantity $F_{\textrm{MCM-CB}}$ that satisfies $F_{\textrm{MCM-CB}} \leq F(\overline{\mathcal{M}})$. 
This is because MCM-CB estimates $F(\overline{\mathcal{M}})$ by averaging the $r_{P,(Z(c_1),Z(c_2))}$, many of which are the geometric mean of multiple $\tilde{\lambda}_{P,(Z(c_1),Z(c_2))}$ [Eq.~\eqref{eq:products}]. 
In the SM, we show that this implies $F_{\textrm{MCM-CB}} \leq F(\overline{\mathcal{M}})$ .

The result of CB of Clifford gates is often used as an estimate of the fidelity of the layer \emph{without} randomized compiling, because randomized compiling of Clifford gates preserves process fidelity (when single-qubit gates are perfect). In contrast, randomized compiling does \emph{not} preserve the fidelity of MCM layers. The fidelity of the randomly compiled instrument is (using Eq~(46) in Ref.~\cite{beale2023randomized})
\begin{align} 
    F(\overline{\mathcal{M}}) & = \Tr(\E_{\substack{G \in \pauli_{n-m}\\}} \E_{k \in \mathbb{Z}_2^m}  (\mathcal{G}^{\dag}\mathcal{V}^{\dag} \otimes \superbra{k})\mathcal{M}_{k} (\mathcal{G} \otimes \superket{k})) \\
    & =   \frac{1}{2^m}\sum_{k \in \mathbb{Z}_2^m}\Tr\left(\mathcal{M}_{k} (\mathcal{V}^{\dag} \otimes \superketbra{k}{k})\right).\label{eq:usi_fidelity}
\end{align}
Using Eq.~\eqref{eq:qi_fidelity}, we conclude that $F(\overline{\mathcal{M}}) \leq F(\mathcal{M})$ for all $\mathcal{M}$.

\begin{figure}[t!]
    \centering
    \includegraphics{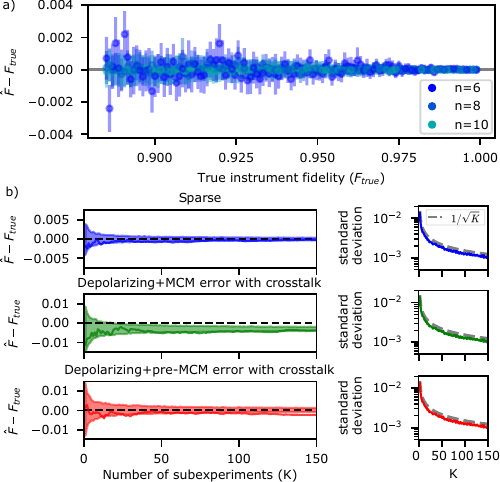}
    \caption{\textbf{Simulating MCM-CB.} (a) MCM-CB of $m=2$ measured qubits and $n-m=4,6,8$ idling qubits with randomly generated USI error models. The MCM-CB estimate is accurate for all error models.  (b) Convergence of the MCM-CB fidelity estimate (solid line, with shaded region showing $\pm1\sigma$ region) with the number of subexperiments ($K$) for three error models with $n=5$ and $m=2$ qubits, computed by sampling $K$ MCM-CB decay parameters $400$ times. The standard deviation of $\hat{F}$ scales with $\nicefrac{1}{\sqrt{K}}$.}
    \label{fig:sim-main}
\end{figure}

\vspace{0.2cm}

\noindent
\textbf{Simulations---}We simulated MCM-CB of layers consisting of MCMs and idle gates, using Stim \cite{Gidney2021stimfaststabilizer}. We randomly sampled USIs of the form $\mathcal{M} = \mathcal{E}_{\textrm{post}}(\mathcal{T} \otimes \mathbb{I}_{n-m})\mathcal{L} \mathcal{E}_{\textrm{pre}}$, where $\mathcal{L}$ is the error-free quantum instrument, $\mathcal{T}$ is a stochastic Pauli channel acting on the unmeasured qubits, and $\mathcal{E}_{\textrm{pre}}$ and $\mathcal{E}_{\textrm{post}}$ are $n$-qubit stochastic Pauli channels where each Pauli error acts nontrivially on at least one measured qubit.  We sampled $3^{n-m}$ uniformly random Pauli error rates for each $\mathcal{E}_{\textrm{pre}}$, $\mathcal{E}_{\textrm{post}}$ and $\mathcal{T}$, normalized to a total error rate that we vary. We also include errors on initial state preparation and final measurement (see SM \cite{supp}). 

We simulated MCM-CB with $K=100$ subexperiments and layers with $n-m=4,6,8$ unmeasured qubits and $m=1,2$ measurements. Figure~\ref{fig:sim-main}a shows the MCM-CB fidelity estimates (all error bars are $1\sigma$ and computed via a standard nonparametric bootstrap). We observe that MCM-CB accurately estimates the USI's fidelity. The uncertainty in the MCM-CB estimates is small for all error models. Most estimates lie within $1\sigma$ of the true fidelity, and all lie within $2.5\sigma$. Figure~\ref{fig:sim-main}c shows how the standard deviation of $\hat{F}$ scales with $K$ for MCM-CB experiments with $n=7$ qubits and $m=2$ measured qubits (details in SM \cite{supp}). The standard deviation decreases with $K$, scaling as $\nicefrac{1}{\sqrt{K}}$ \cite{erhard2019characterizing}.

While MCM-CB accurately estimates fidelity for the error models used in Fig.~\ref{fig:sim-main}a, it only reliably lower bounds the fidelity in general. 
We expect to see larger deviations from the true fidelity with USIs that have low fidelity, and where many of the $\tilde{\lambda}_{P,(Z(c_1),Z(c_2))}$ differ in magnitude from $\tilde{\lambda}_{P,(Z(c_2),Z(c_1))}$ (e.g., because of strong bias towards pre- or post-MCM error). 
The SM presents simulations that explore this effect \cite{supp}.

\begin{figure}[t!]
    \centering
    \includegraphics{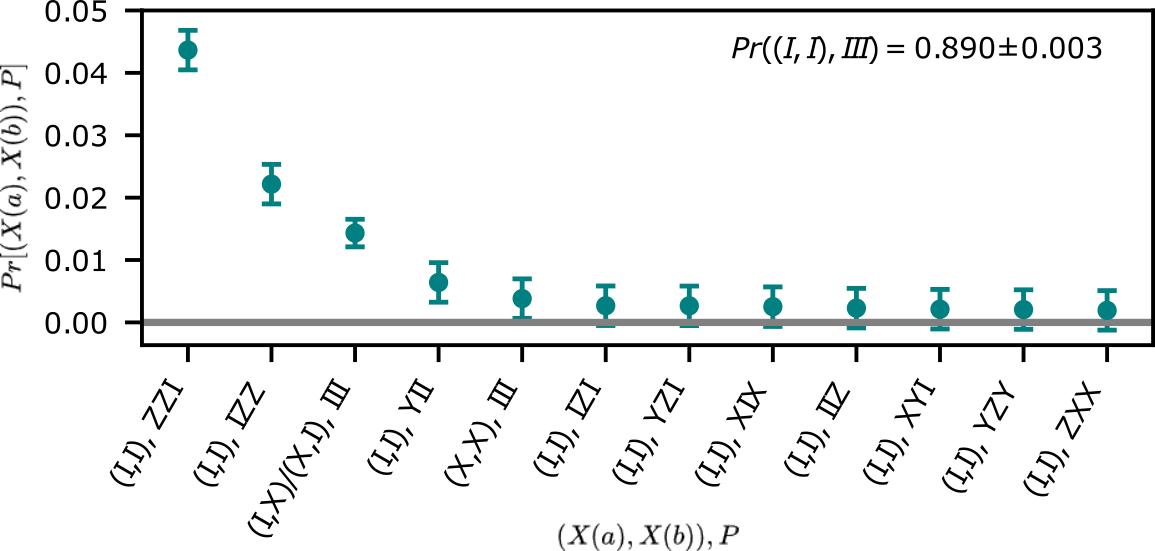}
    \caption{\textbf{MCM-CB on an IBM Q processor.} The $12$ largest Pauli error rates for a $4$-qubit layer of one MCM and three idling qubits on \texttt{ibm\_osaka}. The dominant errors are $ZZ$ errors on adjacent idling qubits and single bit flip errors on the measured qubit.}
    \label{fig:ibmq}
\end{figure}

\vspace{0.2cm}

\noindent
\textbf{IBM Q demonstrations---}We ran MCM-CB on \texttt{ibm\_osaka} to benchmark $n=4$ linearly-connected qubits ($\Q{40}$--$\Q{43}$) with $m=1$ MCMs (on $\Q{40}$), and $n-m=3$ idling qubits with X-X dynamical decoupling. We exhaustively sampled the MCM-CB subexperiments, which required running $64$ sets of MCM-CB circuits and performing $4$ different analyses of each circuit set \footnote{Note that $r_{P,(I,Z)}=r_{P,(Z,I)}$ for the MCM layer we benchmarked, and therefore only 3 analyses are strictly required for each circuit set. We performed both of the equivalent analyses and observed $r_{P,(I,Z)}=r_{P,(Z,I)}$ for all $P$.}. See the SM for full results \cite{supp}. 

Figure~\ref{fig:fig1}c shows the estimated infidelity of the full 4-qubit layer and its subsets (obtained by data marginalization). The errors on the unmeasured subsystem are much larger than those on the measured qubit---the measured qubit's infidelity [$0.022(2)$] is between $1/2$ and $1/4$ of the infidelities of the idling qubits [$0.072(8)$, $0.085(8)$, and $0.040(5)$]. The infidelity of the whole idling subsystem $[0.11(1)]$ is significantly smaller than the sum of the individual idling qubits' infidelities [$0.20(2)$], indicating predominately correlated errors. The infidelity of the idling subsystem $[0.10(1)]$ is within $1\sigma$ of the infidelity of the full $4$-qubit cycle. 

To learn more about the errors, we estimated the Pauli error probabilities for each $\mathcal{T}_{a,b}$. We estimate $\lambda_{0,0,P}$, $\lambda_{1,1,P}$, and $\lambda_{0,1,P}+\lambda_{1,0,P}$ for all $P \in \pauli_3$ and perform a Walsh-Hadamard transform \cite{harper2020efficient} on these three sets of Pauli eigenvalues. 
We use the approximation $\tilde{\lambda}_{P, (I,Z)} \approx \tilde{\lambda}_{P, (Z,I)} \approx\hat{r}_{P, (I,Z)}$ to estimate $\lambda_{0,0,P}$ and $\lambda_{1,1,P}$ (details in SM \cite{supp}). Figure~\ref{fig:ibmq} shows the largest Pauli errors. All but $5$ error rates are within $1\sigma$ of 0. The dominant errors are $ZZ$ errors on adjacent idling qubits [probabilities $0.044(3)$ and $0.022(3)$], consistent with ZZ couplings in transmon qubits \cite{tripathi2022suppression}. The next largest error is single bit flips on the measured qubit [probability $0.014(2)$].

\vspace{0.2cm}

\noindent
\textbf{MCM layer set cycle benchmarking---}MCM-CB of a single layer cannot learn all of the layer's parameters. To learn more parameters, we can benchmark a \emph{set} of layers containing multi-qubit Clifford and MCM layers. To do so, we extend MCM-CB to allow arbitrary single-qubit gates between repetitions of a layer. This means that the learnable parameters are determined by the cycle space of the layer set's \emph{pattern transfer graph (PTG)}, which describes how each layer transforms the support $\supp(P)$ of Pauli operators \cite{chen_learnability_2023}. 
To include MCM layers in the PTG, we add an edge from $\supp(P \otimes Z(c_1))$ to $\supp(P \otimes Z(c_2))$ for each USI PTM element $\tilde{\lambda}_{P, (Z(c_1),Z(c_2))}$. There is an (extended) MCM-CB subexperiment that learns a product of parameters iff it represents a composition of cycles in the PTG. These MCM-CB subexperiments perform a repeated \emph{sequence} of layers. 
For example, by performing MCM-CB subexperiments with the sequence of two-qubit layers consisting of a controlled $Z$ gate followed by a single-qubit MCM, we can learn $12$ parameters that cannot be learned by separately characterizing each layer. These parameters are relational quantities that depend on the error of both layers. MCM-CB is capable of learning every learnable parameter of a layer set, using experiments encoded in the cycles of the PTG (details in SM \cite{supp}).
While we have focused on characterizing individual MCM cycles, interesting applications of our theory include extending Pauli noise learning techniques that benchmark complex circuits or layer sets \cite{harper2020efficient, flammia2022averaged} to those containing MCMs. In particular, we anticipate integrating our method with existing MCM-free techniques for characterizing syndrome extraction circuits \cite{hockings2024scalable, harper2023learning}.

\vspace{0.2cm}
\noindent
\emph{Code implementing MCM-CB is available in pyGSTi \cite{pygstiversion0.9.12.3, nielsen2020probing}.}

\vspace{0.2cm}
\noindent
\emph{Note---After the completion of this work, Ref.~\cite{zhang2024generalized} was posted to arXiv, which introduces a similar method.}

\section*{Acknowledgements} 
This material was funded in part by the U.S. Department of Energy, Office of Science, Office of Advanced Scientific Computing Research, Quantum Testbed Pathfinder Program. T.P. acknowledges support from an Office of Advanced Scientific Computing Research Early Career Award. Sandia National Laboratories is a multi-program laboratory managed and operated by National Technology and Engineering Solutions of Sandia, LLC., a wholly owned subsidiary of Honeywell International, Inc., for the U.S. Department of Energy's National Nuclear Security Administration under contract DE-NA-0003525. This research used IBM Quantum resources of the Air Force Research Laboratory.  All statements of fact, opinion or conclusions contained herein are those of the authors and should not be construed as representing the official views or policies of the U.S. Department of Energy, or the U.S. Government, or IBM, or the IBM Quantum team.

\bibliography{Bibliography}

\begin{thebibliography}{38}%
\makeatletter
\providecommand \@ifxundefined [1]{%
 \@ifx{#1\undefined}
}%
\providecommand \@ifnum [1]{%
 \ifnum #1\expandafter \@firstoftwo
 \else \expandafter \@secondoftwo
 \fi
}%
\providecommand \@ifx [1]{%
 \ifx #1\expandafter \@firstoftwo
 \else \expandafter \@secondoftwo
 \fi
}%
\providecommand \natexlab [1]{#1}%
\providecommand \enquote  [1]{``#1''}%
\providecommand \bibnamefont  [1]{#1}%
\providecommand \bibfnamefont [1]{#1}%
\providecommand \citenamefont [1]{#1}%
\providecommand \href@noop [0]{\@secondoftwo}%
\providecommand \href [0]{\begingroup \@sanitize@url \@href}%
\providecommand \@href[1]{\@@startlink{#1}\@@href}%
\providecommand \@@href[1]{\endgroup#1\@@endlink}%
\providecommand \@sanitize@url [0]{\catcode `\\12\catcode `\$12\catcode `\&12\catcode `\#12\catcode `\^12\catcode `\_12\catcode `\%12\relax}%
\providecommand \@@startlink[1]{}%
\providecommand \@@endlink[0]{}%
\providecommand \url  [0]{\begingroup\@sanitize@url \@url }%
\providecommand \@url [1]{\endgroup\@href {#1}{\urlprefix }}%
\providecommand \urlprefix  [0]{URL }%
\providecommand \Eprint [0]{\href }%
\providecommand \doibase [0]{http://dx.doi.org/}%
\providecommand \selectlanguage [0]{\@gobble}%
\providecommand \bibinfo  [0]{\@secondoftwo}%
\providecommand \bibfield  [0]{\@secondoftwo}%
\providecommand \translation [1]{[#1]}%
\providecommand \BibitemOpen [0]{}%
\providecommand \bibitemStop [0]{}%
\providecommand \bibitemNoStop [0]{.\EOS\space}%
\providecommand \EOS [0]{\spacefactor3000\relax}%
\providecommand \BibitemShut  [1]{\csname bibitem#1\endcsname}%
\let\auto@bib@innerbib\@empty
\bibitem [{\citenamefont {Campbell}\ \emph {et~al.}(2017)\citenamefont {Campbell}, \citenamefont {Terhal},\ and\ \citenamefont {Vuillot}}]{Campbell2017-tw}%
  \BibitemOpen
  \bibfield  {author} {\bibinfo {author} {\bibfnamefont {Earl~T}\ \bibnamefont {Campbell}}, \bibinfo {author} {\bibfnamefont {Barbara~M}\ \bibnamefont {Terhal}}, \ and\ \bibinfo {author} {\bibfnamefont {Christophe}\ \bibnamefont {Vuillot}},\ }\bibfield  {title} {\enquote {\bibinfo {title} {Roads towards fault-tolerant universal quantum computation},}\ }\href {\doibase 10.1038/nature23460} {\bibfield  {journal} {\bibinfo  {journal} {Nature}\ }\textbf {\bibinfo {volume} {549}},\ \bibinfo {pages} {172--179} (\bibinfo {year} {2017})}\BibitemShut {NoStop}%
\bibitem [{\citenamefont {{Google Quantum AI}}(2023)}]{Google_Quantum_AI2023-yr}%
  \BibitemOpen
  \bibfield  {author} {\bibinfo {author} {\bibnamefont {{Google Quantum AI}}},\ }\bibfield  {title} {\enquote {\bibinfo {title} {Suppressing quantum errors by scaling a surface code logical qubit},}\ }\href {\doibase 10.1038/s41586-022-05434-1} {\bibfield  {journal} {\bibinfo  {journal} {Nature}\ }\textbf {\bibinfo {volume} {614}},\ \bibinfo {pages} {676--681} (\bibinfo {year} {2023})}\BibitemShut {NoStop}%
\bibitem [{\citenamefont {Bluvstein}\ \emph {et~al.}(2023)\citenamefont {Bluvstein}, \citenamefont {Evered}, \citenamefont {Geim}, \citenamefont {Li}, \citenamefont {Zhou}, \citenamefont {Manovitz}, \citenamefont {Ebadi}, \citenamefont {Cain}, \citenamefont {Kalinowski}, \citenamefont {Hangleiter}, \citenamefont {Ataides}, \citenamefont {Maskara}, \citenamefont {Cong}, \citenamefont {Gao}, \citenamefont {Rodriguez}, \citenamefont {Karolyshyn}, \citenamefont {Semeghini}, \citenamefont {Gullans}, \citenamefont {Greiner}, \citenamefont {Vuleti{\'c}},\ and\ \citenamefont {Lukin}}]{Bluvstein2023-dp}%
  \BibitemOpen
  \bibfield  {author} {\bibinfo {author} {\bibfnamefont {Dolev}\ \bibnamefont {Bluvstein}}, \bibinfo {author} {\bibfnamefont {Simon~J}\ \bibnamefont {Evered}}, \bibinfo {author} {\bibfnamefont {Alexandra~A}\ \bibnamefont {Geim}}, \bibinfo {author} {\bibfnamefont {Sophie~H}\ \bibnamefont {Li}}, \bibinfo {author} {\bibfnamefont {Hengyun}\ \bibnamefont {Zhou}}, \bibinfo {author} {\bibfnamefont {Tom}\ \bibnamefont {Manovitz}}, \bibinfo {author} {\bibfnamefont {Sepehr}\ \bibnamefont {Ebadi}}, \bibinfo {author} {\bibfnamefont {Madelyn}\ \bibnamefont {Cain}}, \bibinfo {author} {\bibfnamefont {Marcin}\ \bibnamefont {Kalinowski}}, \bibinfo {author} {\bibfnamefont {Dominik}\ \bibnamefont {Hangleiter}}, \bibinfo {author} {\bibfnamefont {J~Pablo~Bonilla}\ \bibnamefont {Ataides}}, \bibinfo {author} {\bibfnamefont {Nishad}\ \bibnamefont {Maskara}}, \bibinfo {author} {\bibfnamefont {Iris}\ \bibnamefont {Cong}}, \bibinfo {author} {\bibfnamefont {Xun}\ \bibnamefont {Gao}}, \bibinfo {author} {\bibfnamefont {Pedro~Sales}\
  \bibnamefont {Rodriguez}}, \bibinfo {author} {\bibfnamefont {Thomas}\ \bibnamefont {Karolyshyn}}, \bibinfo {author} {\bibfnamefont {Giulia}\ \bibnamefont {Semeghini}}, \bibinfo {author} {\bibfnamefont {Michael~J}\ \bibnamefont {Gullans}}, \bibinfo {author} {\bibfnamefont {Markus}\ \bibnamefont {Greiner}}, \bibinfo {author} {\bibfnamefont {Vladan}\ \bibnamefont {Vuleti{\'c}}}, \ and\ \bibinfo {author} {\bibfnamefont {Mikhail~D}\ \bibnamefont {Lukin}},\ }\bibfield  {title} {\enquote {\bibinfo {title} {Logical quantum processor based on reconfigurable atom arrays},}\ }\href {\doibase 10.1038/s41586-023-06927-3} {\bibfield  {journal} {\bibinfo  {journal} {Nature}\ } (\bibinfo {year} {2023}),\ 10.1038/s41586-023-06927-3}\BibitemShut {NoStop}%
\bibitem [{\citenamefont {Krinner}\ \emph {et~al.}(2022)\citenamefont {Krinner}, \citenamefont {Lacroix}, \citenamefont {Remm}, \citenamefont {Di~Paolo}, \citenamefont {Genois}, \citenamefont {Leroux}, \citenamefont {Hellings}, \citenamefont {Lazar}, \citenamefont {Swiadek}, \citenamefont {Herrmann}, \citenamefont {Norris}, \citenamefont {Andersen}, \citenamefont {M{\"u}ller}, \citenamefont {Blais}, \citenamefont {Eichler},\ and\ \citenamefont {Wallraff}}]{Krinner2022-tp}%
  \BibitemOpen
  \bibfield  {author} {\bibinfo {author} {\bibfnamefont {Sebastian}\ \bibnamefont {Krinner}}, \bibinfo {author} {\bibfnamefont {Nathan}\ \bibnamefont {Lacroix}}, \bibinfo {author} {\bibfnamefont {Ants}\ \bibnamefont {Remm}}, \bibinfo {author} {\bibfnamefont {Agustin}\ \bibnamefont {Di~Paolo}}, \bibinfo {author} {\bibfnamefont {Elie}\ \bibnamefont {Genois}}, \bibinfo {author} {\bibfnamefont {Catherine}\ \bibnamefont {Leroux}}, \bibinfo {author} {\bibfnamefont {Christoph}\ \bibnamefont {Hellings}}, \bibinfo {author} {\bibfnamefont {Stefania}\ \bibnamefont {Lazar}}, \bibinfo {author} {\bibfnamefont {Francois}\ \bibnamefont {Swiadek}}, \bibinfo {author} {\bibfnamefont {Johannes}\ \bibnamefont {Herrmann}}, \bibinfo {author} {\bibfnamefont {Graham~J}\ \bibnamefont {Norris}}, \bibinfo {author} {\bibfnamefont {Christian~Kraglund}\ \bibnamefont {Andersen}}, \bibinfo {author} {\bibfnamefont {Markus}\ \bibnamefont {M{\"u}ller}}, \bibinfo {author} {\bibfnamefont {Alexandre}\ \bibnamefont {Blais}}, \bibinfo {author}
  {\bibfnamefont {Christopher}\ \bibnamefont {Eichler}}, \ and\ \bibinfo {author} {\bibfnamefont {Andreas}\ \bibnamefont {Wallraff}},\ }\bibfield  {title} {\enquote {\bibinfo {title} {Realizing repeated quantum error correction in a distance-three surface code},}\ }\href {\doibase 10.1038/s41586-022-04566-8} {\bibfield  {journal} {\bibinfo  {journal} {Nature}\ }\textbf {\bibinfo {volume} {605}},\ \bibinfo {pages} {669--674} (\bibinfo {year} {2022})}\BibitemShut {NoStop}%
\bibitem [{\citenamefont {Gupta}\ \emph {et~al.}(2024{\natexlab{a}})\citenamefont {Gupta}, \citenamefont {Sundaresan}, \citenamefont {Alexander}, \citenamefont {Wood}, \citenamefont {Merkel}, \citenamefont {Healy}, \citenamefont {Hillenbrand}, \citenamefont {Jochym-O'Connor}, \citenamefont {Wootton}, \citenamefont {Yoder}, \citenamefont {Cross}, \citenamefont {Takita},\ and\ \citenamefont {Brown}}]{Gupta2024-pr}%
  \BibitemOpen
  \bibfield  {author} {\bibinfo {author} {\bibfnamefont {Riddhi~S}\ \bibnamefont {Gupta}}, \bibinfo {author} {\bibfnamefont {Neereja}\ \bibnamefont {Sundaresan}}, \bibinfo {author} {\bibfnamefont {Thomas}\ \bibnamefont {Alexander}}, \bibinfo {author} {\bibfnamefont {Christopher~J}\ \bibnamefont {Wood}}, \bibinfo {author} {\bibfnamefont {Seth~T}\ \bibnamefont {Merkel}}, \bibinfo {author} {\bibfnamefont {Michael~B}\ \bibnamefont {Healy}}, \bibinfo {author} {\bibfnamefont {Marius}\ \bibnamefont {Hillenbrand}}, \bibinfo {author} {\bibfnamefont {Tomas}\ \bibnamefont {Jochym-O'Connor}}, \bibinfo {author} {\bibfnamefont {James~R}\ \bibnamefont {Wootton}}, \bibinfo {author} {\bibfnamefont {Theodore~J}\ \bibnamefont {Yoder}}, \bibinfo {author} {\bibfnamefont {Andrew~W}\ \bibnamefont {Cross}}, \bibinfo {author} {\bibfnamefont {Maika}\ \bibnamefont {Takita}}, \ and\ \bibinfo {author} {\bibfnamefont {Benjamin~J}\ \bibnamefont {Brown}},\ }\bibfield  {title} {\enquote {\bibinfo {title} {Encoding a magic state with beyond
  break-even fidelity},}\ }\href {\doibase 10.1038/s41586-023-06846-3} {\bibfield  {journal} {\bibinfo  {journal} {Nature}\ }\textbf {\bibinfo {volume} {625}},\ \bibinfo {pages} {259--263} (\bibinfo {year} {2024}{\natexlab{a}})}\BibitemShut {NoStop}%
\bibitem [{\citenamefont {Lu}\ \emph {et~al.}(2022)\citenamefont {Lu}, \citenamefont {Lessa}, \citenamefont {Kim},\ and\ \citenamefont {Hsieh}}]{lu2022measurement}%
  \BibitemOpen
  \bibfield  {author} {\bibinfo {author} {\bibfnamefont {Tsung-Cheng}\ \bibnamefont {Lu}}, \bibinfo {author} {\bibfnamefont {Leonardo~A.}\ \bibnamefont {Lessa}}, \bibinfo {author} {\bibfnamefont {Isaac~H.}\ \bibnamefont {Kim}}, \ and\ \bibinfo {author} {\bibfnamefont {Timothy~H.}\ \bibnamefont {Hsieh}},\ }\bibfield  {title} {\enquote {\bibinfo {title} {Measurement as a shortcut to long-range entangled quantum matter},}\ }\href {\doibase 10.1103/PRXQuantum.3.040337} {\bibfield  {journal} {\bibinfo  {journal} {PRX Quantum}\ }\textbf {\bibinfo {volume} {3}},\ \bibinfo {pages} {040337} (\bibinfo {year} {2022})}\BibitemShut {NoStop}%
\bibitem [{\citenamefont {Griffiths}\ and\ \citenamefont {Niu}(1996)}]{griffiths1996semiclassical}%
  \BibitemOpen
  \bibfield  {author} {\bibinfo {author} {\bibfnamefont {Robert~B.}\ \bibnamefont {Griffiths}}\ and\ \bibinfo {author} {\bibfnamefont {Chi-Sheng}\ \bibnamefont {Niu}},\ }\bibfield  {title} {\enquote {\bibinfo {title} {Semiclassical fourier transform for quantum computation},}\ }\href {\doibase 10.1103/PhysRevLett.76.3228} {\bibfield  {journal} {\bibinfo  {journal} {Phys. Rev. Lett.}\ }\textbf {\bibinfo {volume} {76}},\ \bibinfo {pages} {3228--3231} (\bibinfo {year} {1996})}\BibitemShut {NoStop}%
\bibitem [{\citenamefont {Bäumer}\ \emph {et~al.}(2023)\citenamefont {Bäumer}, \citenamefont {Tripathi}, \citenamefont {Wang}, \citenamefont {Rall}, \citenamefont {Chen}, \citenamefont {Majumder}, \citenamefont {Seif},\ and\ \citenamefont {Minev}}]{bäumer2023efficient}%
  \BibitemOpen
  \bibfield  {author} {\bibinfo {author} {\bibfnamefont {Elisa}\ \bibnamefont {Bäumer}}, \bibinfo {author} {\bibfnamefont {Vinay}\ \bibnamefont {Tripathi}}, \bibinfo {author} {\bibfnamefont {Derek~S.}\ \bibnamefont {Wang}}, \bibinfo {author} {\bibfnamefont {Patrick}\ \bibnamefont {Rall}}, \bibinfo {author} {\bibfnamefont {Edward~H.}\ \bibnamefont {Chen}}, \bibinfo {author} {\bibfnamefont {Swarnadeep}\ \bibnamefont {Majumder}}, \bibinfo {author} {\bibfnamefont {Alireza}\ \bibnamefont {Seif}}, \ and\ \bibinfo {author} {\bibfnamefont {Zlatko~K.}\ \bibnamefont {Minev}},\ }\href@noop {} {\enquote {\bibinfo {title} {Efficient long-range entanglement using dynamic circuits},}\ } (\bibinfo {year} {2023}),\ \Eprint {http://arxiv.org/abs/2308.13065} {arXiv:2308.13065 [quant-ph]} \BibitemShut {NoStop}%
\bibitem [{\citenamefont {Rudinger}\ \emph {et~al.}(2022)\citenamefont {Rudinger}, \citenamefont {Ribeill}, \citenamefont {Govia}, \citenamefont {Ware}, \citenamefont {Nielsen}, \citenamefont {Young}, \citenamefont {Ohki}, \citenamefont {Blume-Kohout},\ and\ \citenamefont {Proctor}}]{rudinger2022characterizing}%
  \BibitemOpen
  \bibfield  {author} {\bibinfo {author} {\bibfnamefont {Kenneth}\ \bibnamefont {Rudinger}}, \bibinfo {author} {\bibfnamefont {Guilhem~J.}\ \bibnamefont {Ribeill}}, \bibinfo {author} {\bibfnamefont {Luke~C.G.}\ \bibnamefont {Govia}}, \bibinfo {author} {\bibfnamefont {Matthew}\ \bibnamefont {Ware}}, \bibinfo {author} {\bibfnamefont {Erik}\ \bibnamefont {Nielsen}}, \bibinfo {author} {\bibfnamefont {Kevin}\ \bibnamefont {Young}}, \bibinfo {author} {\bibfnamefont {Thomas~A.}\ \bibnamefont {Ohki}}, \bibinfo {author} {\bibfnamefont {Robin}\ \bibnamefont {Blume-Kohout}}, \ and\ \bibinfo {author} {\bibfnamefont {Timothy}\ \bibnamefont {Proctor}},\ }\bibfield  {title} {\enquote {\bibinfo {title} {Characterizing midcircuit measurements on a superconducting qubit using gate set tomography},}\ }\href {\doibase 10.1103/PhysRevApplied.17.014014} {\bibfield  {journal} {\bibinfo  {journal} {Phys. Rev. Appl.}\ }\textbf {\bibinfo {volume} {17}},\ \bibinfo {pages} {014014} (\bibinfo {year} {2022})}\BibitemShut {NoStop}%
\bibitem [{\citenamefont {Stricker}\ \emph {et~al.}(2022)\citenamefont {Stricker}, \citenamefont {Vodola}, \citenamefont {Erhard}, \citenamefont {Postler}, \citenamefont {Meth}, \citenamefont {Ringbauer}, \citenamefont {Schindler}, \citenamefont {Blatt}, \citenamefont {M\"uller},\ and\ \citenamefont {Monz}}]{strickler2022characterizing}%
  \BibitemOpen
  \bibfield  {author} {\bibinfo {author} {\bibfnamefont {Roman}\ \bibnamefont {Stricker}}, \bibinfo {author} {\bibfnamefont {Davide}\ \bibnamefont {Vodola}}, \bibinfo {author} {\bibfnamefont {Alexander}\ \bibnamefont {Erhard}}, \bibinfo {author} {\bibfnamefont {Lukas}\ \bibnamefont {Postler}}, \bibinfo {author} {\bibfnamefont {Michael}\ \bibnamefont {Meth}}, \bibinfo {author} {\bibfnamefont {Martin}\ \bibnamefont {Ringbauer}}, \bibinfo {author} {\bibfnamefont {Philipp}\ \bibnamefont {Schindler}}, \bibinfo {author} {\bibfnamefont {Rainer}\ \bibnamefont {Blatt}}, \bibinfo {author} {\bibfnamefont {Markus}\ \bibnamefont {M\"uller}}, \ and\ \bibinfo {author} {\bibfnamefont {Thomas}\ \bibnamefont {Monz}},\ }\bibfield  {title} {\enquote {\bibinfo {title} {Characterizing quantum instruments: From nondemolition measurements to quantum error correction},}\ }\href {\doibase 10.1103/PRXQuantum.3.030318} {\bibfield  {journal} {\bibinfo  {journal} {PRX Quantum}\ }\textbf {\bibinfo {volume} {3}},\ \bibinfo {pages} {030318}
  (\bibinfo {year} {2022})}\BibitemShut {NoStop}%
\bibitem [{\citenamefont {Govia}\ \emph {et~al.}(2023)\citenamefont {Govia}, \citenamefont {Jurcevic}, \citenamefont {Wood}, \citenamefont {Kanazawa}, \citenamefont {Merkel},\ and\ \citenamefont {McKay}}]{govia2023randomized}%
  \BibitemOpen
  \bibfield  {author} {\bibinfo {author} {\bibfnamefont {L~C~G}\ \bibnamefont {Govia}}, \bibinfo {author} {\bibfnamefont {P}~\bibnamefont {Jurcevic}}, \bibinfo {author} {\bibfnamefont {C~J}\ \bibnamefont {Wood}}, \bibinfo {author} {\bibfnamefont {N}~\bibnamefont {Kanazawa}}, \bibinfo {author} {\bibfnamefont {S~T}\ \bibnamefont {Merkel}}, \ and\ \bibinfo {author} {\bibfnamefont {D~C}\ \bibnamefont {McKay}},\ }\bibfield  {title} {\enquote {\bibinfo {title} {A randomized benchmarking suite for mid-circuit measurements},}\ }\href {\doibase 10.1088/1367-2630/ad0e19} {\bibfield  {journal} {\bibinfo  {journal} {New Journal of Physics}\ }\textbf {\bibinfo {volume} {25}},\ \bibinfo {pages} {123016} (\bibinfo {year} {2023})}\BibitemShut {NoStop}%
\bibitem [{\citenamefont {Gaebler}\ \emph {et~al.}(2021)\citenamefont {Gaebler}, \citenamefont {Baldwin}, \citenamefont {Moses}, \citenamefont {Dreiling}, \citenamefont {Figgatt}, \citenamefont {Foss-Feig}, \citenamefont {Hayes},\ and\ \citenamefont {Pino}}]{gaebler2021suppression}%
  \BibitemOpen
  \bibfield  {author} {\bibinfo {author} {\bibfnamefont {J.~P.}\ \bibnamefont {Gaebler}}, \bibinfo {author} {\bibfnamefont {C.~H.}\ \bibnamefont {Baldwin}}, \bibinfo {author} {\bibfnamefont {S.~A.}\ \bibnamefont {Moses}}, \bibinfo {author} {\bibfnamefont {J.~M.}\ \bibnamefont {Dreiling}}, \bibinfo {author} {\bibfnamefont {C.}~\bibnamefont {Figgatt}}, \bibinfo {author} {\bibfnamefont {M.}~\bibnamefont {Foss-Feig}}, \bibinfo {author} {\bibfnamefont {D.}~\bibnamefont {Hayes}}, \ and\ \bibinfo {author} {\bibfnamefont {J.~M.}\ \bibnamefont {Pino}},\ }\bibfield  {title} {\enquote {\bibinfo {title} {Suppression of midcircuit measurement crosstalk errors with micromotion},}\ }\href {\doibase 10.1103/PhysRevA.104.062440} {\bibfield  {journal} {\bibinfo  {journal} {Phys. Rev. A}\ }\textbf {\bibinfo {volume} {104}},\ \bibinfo {pages} {062440} (\bibinfo {year} {2021})}\BibitemShut {NoStop}%
\bibitem [{\citenamefont {Flammia}\ and\ \citenamefont {Wallman}(2020)}]{flammia2020efficient}%
  \BibitemOpen
  \bibfield  {author} {\bibinfo {author} {\bibfnamefont {Steven~T.}\ \bibnamefont {Flammia}}\ and\ \bibinfo {author} {\bibfnamefont {Joel~J.}\ \bibnamefont {Wallman}},\ }\bibfield  {title} {\enquote {\bibinfo {title} {Efficient estimation of pauli channels},}\ }\href {\doibase 10.1145/3408039} {\bibfield  {journal} {\bibinfo  {journal} {ACM Transactions on Quantum Computing}\ }\textbf {\bibinfo {volume} {1}},\ \bibinfo {pages} {1–32} (\bibinfo {year} {2020})}\BibitemShut {NoStop}%
\bibitem [{\citenamefont {Harper}\ \emph {et~al.}(2020)\citenamefont {Harper}, \citenamefont {Flammia},\ and\ \citenamefont {Wallman}}]{harper2020efficient}%
  \BibitemOpen
  \bibfield  {author} {\bibinfo {author} {\bibfnamefont {Robin}\ \bibnamefont {Harper}}, \bibinfo {author} {\bibfnamefont {Steven~T.}\ \bibnamefont {Flammia}}, \ and\ \bibinfo {author} {\bibfnamefont {Joel~J.}\ \bibnamefont {Wallman}},\ }\bibfield  {title} {\enquote {\bibinfo {title} {Efficient learning of quantum noise},}\ }\href {\doibase 10.1038/s41567-020-0992-8} {\bibfield  {journal} {\bibinfo  {journal} {Nature Physics}\ }\textbf {\bibinfo {volume} {16}},\ \bibinfo {pages} {1184–1188} (\bibinfo {year} {2020})}\BibitemShut {NoStop}%
\bibitem [{\citenamefont {Harper}\ \emph {et~al.}(2021)\citenamefont {Harper}, \citenamefont {Yu},\ and\ \citenamefont {Flammia}}]{harper2021fast}%
  \BibitemOpen
  \bibfield  {author} {\bibinfo {author} {\bibfnamefont {Robin}\ \bibnamefont {Harper}}, \bibinfo {author} {\bibfnamefont {Wenjun}\ \bibnamefont {Yu}}, \ and\ \bibinfo {author} {\bibfnamefont {Steven~T.}\ \bibnamefont {Flammia}},\ }\bibfield  {title} {\enquote {\bibinfo {title} {Fast estimation of sparse quantum noise},}\ }\href {\doibase 10.1103/PRXQuantum.2.010322} {\bibfield  {journal} {\bibinfo  {journal} {PRX Quantum}\ }\textbf {\bibinfo {volume} {2}},\ \bibinfo {pages} {010322} (\bibinfo {year} {2021})}\BibitemShut {NoStop}%
\bibitem [{\citenamefont {Chen}\ \emph {et~al.}(2023)\citenamefont {Chen}, \citenamefont {Liu}, \citenamefont {Otten}, \citenamefont {Seif}, \citenamefont {Fefferman},\ and\ \citenamefont {Jiang}}]{chen_learnability_2023}%
  \BibitemOpen
  \bibfield  {author} {\bibinfo {author} {\bibfnamefont {Senrui}\ \bibnamefont {Chen}}, \bibinfo {author} {\bibfnamefont {Yunchao}\ \bibnamefont {Liu}}, \bibinfo {author} {\bibfnamefont {Matthew}\ \bibnamefont {Otten}}, \bibinfo {author} {\bibfnamefont {Alireza}\ \bibnamefont {Seif}}, \bibinfo {author} {\bibfnamefont {Bill}\ \bibnamefont {Fefferman}}, \ and\ \bibinfo {author} {\bibfnamefont {Liang}\ \bibnamefont {Jiang}},\ }\bibfield  {title} {\enquote {\bibinfo {title} {The learnability of {Pauli} noise},}\ }\href {\doibase 10.1038/s41467-022-35759-4} {\bibfield  {journal} {\bibinfo  {journal} {Nature Communications}\ }\textbf {\bibinfo {volume} {14}},\ \bibinfo {pages} {52} (\bibinfo {year} {2023})},\ \bibinfo {note} {publisher: Nature Publishing Group}\BibitemShut {NoStop}%
\bibitem [{\citenamefont {Flammia}(2022)}]{flammia2022averaged}%
  \BibitemOpen
  \bibfield  {author} {\bibinfo {author} {\bibfnamefont {Steven~T.}\ \bibnamefont {Flammia}},\ }\bibfield  {title} {\enquote {\bibinfo {title} {Averaged circuit eigenvalue sampling},}\ \ }(\bibinfo  {publisher} {Schloss Dagstuhl – Leibniz-Zentrum für Informatik},\ \bibinfo {year} {2022})\BibitemShut {NoStop}%
\bibitem [{\citenamefont {Temme}\ \emph {et~al.}(2017)\citenamefont {Temme}, \citenamefont {Bravyi},\ and\ \citenamefont {Gambetta}}]{temme2017error}%
  \BibitemOpen
  \bibfield  {author} {\bibinfo {author} {\bibfnamefont {Kristan}\ \bibnamefont {Temme}}, \bibinfo {author} {\bibfnamefont {Sergey}\ \bibnamefont {Bravyi}}, \ and\ \bibinfo {author} {\bibfnamefont {Jay~M.}\ \bibnamefont {Gambetta}},\ }\bibfield  {title} {\enquote {\bibinfo {title} {Error mitigation for short-depth quantum circuits},}\ }\href {\doibase 10.1103/PhysRevLett.119.180509} {\bibfield  {journal} {\bibinfo  {journal} {Phys. Rev. Lett.}\ }\textbf {\bibinfo {volume} {119}},\ \bibinfo {pages} {180509} (\bibinfo {year} {2017})}\BibitemShut {NoStop}%
\bibitem [{\citenamefont {van~den Berg}\ \emph {et~al.}(2023)\citenamefont {van~den Berg}, \citenamefont {Minev}, \citenamefont {Kandala},\ and\ \citenamefont {Temme}}]{vandenBerg2023}%
  \BibitemOpen
  \bibfield  {author} {\bibinfo {author} {\bibfnamefont {Ewout}\ \bibnamefont {van~den Berg}}, \bibinfo {author} {\bibfnamefont {Zlatko~K.}\ \bibnamefont {Minev}}, \bibinfo {author} {\bibfnamefont {Abhinav}\ \bibnamefont {Kandala}}, \ and\ \bibinfo {author} {\bibfnamefont {Kristan}\ \bibnamefont {Temme}},\ }\bibfield  {title} {\enquote {\bibinfo {title} {Probabilistic error cancellation with sparse pauli--lindblad models on noisy quantum processors},}\ }\href {\doibase 10.1038/s41567-023-02042-2} {\bibfield  {journal} {\bibinfo  {journal} {Nature Physics}\ }\textbf {\bibinfo {volume} {19}},\ \bibinfo {pages} {1116--1121} (\bibinfo {year} {2023})}\BibitemShut {NoStop}%
\bibitem [{\citenamefont {Gupta}\ \emph {et~al.}(2024{\natexlab{b}})\citenamefont {Gupta}, \citenamefont {van~den Berg}, \citenamefont {Takita}, \citenamefont {Rist\`e}, \citenamefont {Temme},\ and\ \citenamefont {Kandala}}]{gupta_probabilistic_2023}%
  \BibitemOpen
  \bibfield  {author} {\bibinfo {author} {\bibfnamefont {Riddhi~S.}\ \bibnamefont {Gupta}}, \bibinfo {author} {\bibfnamefont {Ewout}\ \bibnamefont {van~den Berg}}, \bibinfo {author} {\bibfnamefont {Maika}\ \bibnamefont {Takita}}, \bibinfo {author} {\bibfnamefont {Diego}\ \bibnamefont {Rist\`e}}, \bibinfo {author} {\bibfnamefont {Kristan}\ \bibnamefont {Temme}}, \ and\ \bibinfo {author} {\bibfnamefont {Abhinav}\ \bibnamefont {Kandala}},\ }\bibfield  {title} {\enquote {\bibinfo {title} {Probabilistic error cancellation for dynamic quantum circuits},}\ }\href {\doibase 10.1103/PhysRevA.109.062617} {\bibfield  {journal} {\bibinfo  {journal} {Phys. Rev. A}\ }\textbf {\bibinfo {volume} {109}},\ \bibinfo {pages} {062617} (\bibinfo {year} {2024}{\natexlab{b}})}\BibitemShut {NoStop}%
\bibitem [{\citenamefont {Harper}\ and\ \citenamefont {Flammia}(2023)}]{harper2023learning}%
  \BibitemOpen
  \bibfield  {author} {\bibinfo {author} {\bibfnamefont {Robin}\ \bibnamefont {Harper}}\ and\ \bibinfo {author} {\bibfnamefont {Steven~T.}\ \bibnamefont {Flammia}},\ }\bibfield  {title} {\enquote {\bibinfo {title} {Learning correlated noise in a 39-qubit quantum processor},}\ }\href {\doibase 10.1103/PRXQuantum.4.040311} {\bibfield  {journal} {\bibinfo  {journal} {PRX Quantum}\ }\textbf {\bibinfo {volume} {4}},\ \bibinfo {pages} {040311} (\bibinfo {year} {2023})}\BibitemShut {NoStop}%
\bibitem [{\citenamefont {Hockings}\ \emph {et~al.}(2024)\citenamefont {Hockings}, \citenamefont {Doherty},\ and\ \citenamefont {Harper}}]{hockings2024scalable}%
  \BibitemOpen
  \bibfield  {author} {\bibinfo {author} {\bibfnamefont {Evan~T.}\ \bibnamefont {Hockings}}, \bibinfo {author} {\bibfnamefont {Andrew~C.}\ \bibnamefont {Doherty}}, \ and\ \bibinfo {author} {\bibfnamefont {Robin}\ \bibnamefont {Harper}},\ }\href@noop {} {\enquote {\bibinfo {title} {Scalable noise characterisation of syndrome extraction circuits with averaged circuit eigenvalue sampling},}\ } (\bibinfo {year} {2024}),\ \Eprint {http://arxiv.org/abs/2404.06545} {arXiv:2404.06545 [quant-ph]} \BibitemShut {NoStop}%
\bibitem [{\citenamefont {Wallman}\ and\ \citenamefont {Emerson}(2016)}]{wallman2016noise}%
  \BibitemOpen
  \bibfield  {author} {\bibinfo {author} {\bibfnamefont {Joel~J.}\ \bibnamefont {Wallman}}\ and\ \bibinfo {author} {\bibfnamefont {Joseph}\ \bibnamefont {Emerson}},\ }\bibfield  {title} {\enquote {\bibinfo {title} {Noise tailoring for scalable quantum computation via randomized compiling},}\ }\href {\doibase 10.1103/PhysRevA.94.052325} {\bibfield  {journal} {\bibinfo  {journal} {Phys. Rev. A}\ }\textbf {\bibinfo {volume} {94}},\ \bibinfo {pages} {052325} (\bibinfo {year} {2016})}\BibitemShut {NoStop}%
\bibitem [{\citenamefont {Beale}\ and\ \citenamefont {Wallman}(2023)}]{beale2023randomized}%
  \BibitemOpen
  \bibfield  {author} {\bibinfo {author} {\bibfnamefont {Stefanie~J.}\ \bibnamefont {Beale}}\ and\ \bibinfo {author} {\bibfnamefont {Joel~J.}\ \bibnamefont {Wallman}},\ }\href@noop {} {\enquote {\bibinfo {title} {Randomized compiling for subsystem measurements},}\ } (\bibinfo {year} {2023}),\ \Eprint {http://arxiv.org/abs/2304.06599} {arXiv:2304.06599 [quant-ph]} \BibitemShut {NoStop}%
\bibitem [{\citenamefont {Hashim}\ \emph {et~al.}(2024)\citenamefont {Hashim}, \citenamefont {Carignan-Dugas}, \citenamefont {Chen}, \citenamefont {Juenger}, \citenamefont {Fruitwala}, \citenamefont {Xu}, \citenamefont {Huang}, \citenamefont {Wallman},\ and\ \citenamefont {Siddiqi}}]{hashim2024quasiprobabilistic}%
  \BibitemOpen
  \bibfield  {author} {\bibinfo {author} {\bibfnamefont {Akel}\ \bibnamefont {Hashim}}, \bibinfo {author} {\bibfnamefont {Arnaud}\ \bibnamefont {Carignan-Dugas}}, \bibinfo {author} {\bibfnamefont {Larry}\ \bibnamefont {Chen}}, \bibinfo {author} {\bibfnamefont {Christian}\ \bibnamefont {Juenger}}, \bibinfo {author} {\bibfnamefont {Neelay}\ \bibnamefont {Fruitwala}}, \bibinfo {author} {\bibfnamefont {Yilun}\ \bibnamefont {Xu}}, \bibinfo {author} {\bibfnamefont {Gang}\ \bibnamefont {Huang}}, \bibinfo {author} {\bibfnamefont {Joel~J.}\ \bibnamefont {Wallman}}, \ and\ \bibinfo {author} {\bibfnamefont {Irfan}\ \bibnamefont {Siddiqi}},\ }\href@noop {} {\enquote {\bibinfo {title} {Quasi-probabilistic readout correction of mid-circuit measurements for adaptive feedback via measurement randomized compiling},}\ } (\bibinfo {year} {2024}),\ \Eprint {http://arxiv.org/abs/2312.14139} {arXiv:2312.14139 [quant-ph]} \BibitemShut {NoStop}%
\bibitem [{\citenamefont {McLaren}\ \emph {et~al.}(2023)\citenamefont {McLaren}, \citenamefont {Graydon},\ and\ \citenamefont {Wallman}}]{mclaren2023stochastic}%
  \BibitemOpen
  \bibfield  {author} {\bibinfo {author} {\bibfnamefont {Darian}\ \bibnamefont {McLaren}}, \bibinfo {author} {\bibfnamefont {Matthew~A.}\ \bibnamefont {Graydon}}, \ and\ \bibinfo {author} {\bibfnamefont {Joel~J.}\ \bibnamefont {Wallman}},\ }\href@noop {} {\enquote {\bibinfo {title} {Stochastic errors in quantum instruments},}\ } (\bibinfo {year} {2023}),\ \Eprint {http://arxiv.org/abs/2306.07418} {arXiv:2306.07418 [quant-ph]} \BibitemShut {NoStop}%
\bibitem [{\citenamefont {Erhard}\ \emph {et~al.}(2019)\citenamefont {Erhard}, \citenamefont {Wallman}, \citenamefont {Postler}, \citenamefont {Meth}, \citenamefont {Stricker}, \citenamefont {Martinez}, \citenamefont {Schindler}, \citenamefont {Monz}, \citenamefont {Emerson},\ and\ \citenamefont {Blatt}}]{erhard2019characterizing}%
  \BibitemOpen
  \bibfield  {author} {\bibinfo {author} {\bibfnamefont {Alexander}\ \bibnamefont {Erhard}}, \bibinfo {author} {\bibfnamefont {Joel~J.}\ \bibnamefont {Wallman}}, \bibinfo {author} {\bibfnamefont {Lukas}\ \bibnamefont {Postler}}, \bibinfo {author} {\bibfnamefont {Michael}\ \bibnamefont {Meth}}, \bibinfo {author} {\bibfnamefont {Roman}\ \bibnamefont {Stricker}}, \bibinfo {author} {\bibfnamefont {Esteban~A.}\ \bibnamefont {Martinez}}, \bibinfo {author} {\bibfnamefont {Philipp}\ \bibnamefont {Schindler}}, \bibinfo {author} {\bibfnamefont {Thomas}\ \bibnamefont {Monz}}, \bibinfo {author} {\bibfnamefont {Joseph}\ \bibnamefont {Emerson}}, \ and\ \bibinfo {author} {\bibfnamefont {Rainer}\ \bibnamefont {Blatt}},\ }\bibfield  {title} {\enquote {\bibinfo {title} {Characterizing large-scale quantum computers via cycle benchmarking},}\ }\href {\doibase 10.1038/s41467-019-13068-7} {\bibfield  {journal} {\bibinfo  {journal} {Nature Communications}\ }\textbf {\bibinfo {volume} {10}} (\bibinfo {year} {2019}),\
  10.1038/s41467-019-13068-7}\BibitemShut {NoStop}%
\bibitem [{\citenamefont {Carignan-Dugas}\ \emph {et~al.}(2023)\citenamefont {Carignan-Dugas}, \citenamefont {Ranu},\ and\ \citenamefont {Dreher}}]{carignan-dugas_estimating_2023}%
  \BibitemOpen
  \bibfield  {author} {\bibinfo {author} {\bibfnamefont {Arnaud}\ \bibnamefont {Carignan-Dugas}}, \bibinfo {author} {\bibfnamefont {Shashank~Kumar}\ \bibnamefont {Ranu}}, \ and\ \bibinfo {author} {\bibfnamefont {Patrick}\ \bibnamefont {Dreher}},\ }\href {http://arxiv.org/abs/2303.09945} {\enquote {\bibinfo {title} {Estimating {Coherent} {Contributions} to the {Error} {Profile} {Using} {Cycle} {Error} {Reconstruction}},}\ } (\bibinfo {year} {2023}),\ \bibinfo {note} {arXiv:2303.09945 [quant-ph]}\BibitemShut {NoStop}%
\bibitem [{sup()}]{supp}%
  \BibitemOpen
  \href@noop {} {}\bibinfo {note} {See Supplemental Material at [URL-will-be-inserted-by-publisher] for further details of the theory and additional results from our simulations and IBM Q demonstrations.}\BibitemShut {Stop}%
\bibitem [{Note1()}]{Note1}%
  \BibitemOpen
  \bibinfo {note} {Eq.\protect \eqref {eq:lambda_tilde} is a binary Fourier transform, and therefore $\protect \tilde {\lambda }_{P,(Z(c), Z(c'))}$ can also be thought of as a Fourier coefficient \cite {zhang2024generalized}.}\BibitemShut {Stop}%
\bibitem [{\citenamefont {Kim}\ \emph {et~al.}(2022)\citenamefont {Kim}, \citenamefont {Morvan}, \citenamefont {Nguyen}, \citenamefont {Naik}, \citenamefont {J{\"u}nger}, \citenamefont {Chen}, \citenamefont {Kreikebaum}, \citenamefont {Santiago},\ and\ \citenamefont {Siddiqi}}]{Kim2022}%
  \BibitemOpen
  \bibfield  {author} {\bibinfo {author} {\bibfnamefont {Yosep}\ \bibnamefont {Kim}}, \bibinfo {author} {\bibfnamefont {Alexis}\ \bibnamefont {Morvan}}, \bibinfo {author} {\bibfnamefont {Long~B.}\ \bibnamefont {Nguyen}}, \bibinfo {author} {\bibfnamefont {Ravi~K.}\ \bibnamefont {Naik}}, \bibinfo {author} {\bibfnamefont {Christian}\ \bibnamefont {J{\"u}nger}}, \bibinfo {author} {\bibfnamefont {Larry}\ \bibnamefont {Chen}}, \bibinfo {author} {\bibfnamefont {John~Mark}\ \bibnamefont {Kreikebaum}}, \bibinfo {author} {\bibfnamefont {David~I.}\ \bibnamefont {Santiago}}, \ and\ \bibinfo {author} {\bibfnamefont {Irfan}\ \bibnamefont {Siddiqi}},\ }\bibfield  {title} {\enquote {\bibinfo {title} {High-fidelity three-qubit itoffoli gate for fixed-frequency superconducting qubits},}\ }\href {\doibase 10.1038/s41567-022-01590-3} {\bibfield  {journal} {\bibinfo  {journal} {Nature Physics}\ }\textbf {\bibinfo {volume} {18}},\ \bibinfo {pages} {783--788} (\bibinfo {year} {2022})}\BibitemShut {NoStop}%
\bibitem [{\citenamefont {Hashim}\ \emph {et~al.}(2022)\citenamefont {Hashim}, \citenamefont {Rines}, \citenamefont {Omole}, \citenamefont {Naik}, \citenamefont {Kreikebaum}, \citenamefont {Santiago}, \citenamefont {Chong}, \citenamefont {Siddiqi},\ and\ \citenamefont {Gokhale}}]{hashim2022optimized}%
  \BibitemOpen
  \bibfield  {author} {\bibinfo {author} {\bibfnamefont {Akel}\ \bibnamefont {Hashim}}, \bibinfo {author} {\bibfnamefont {Rich}\ \bibnamefont {Rines}}, \bibinfo {author} {\bibfnamefont {Victory}\ \bibnamefont {Omole}}, \bibinfo {author} {\bibfnamefont {Ravi~K.}\ \bibnamefont {Naik}}, \bibinfo {author} {\bibfnamefont {John~Mark}\ \bibnamefont {Kreikebaum}}, \bibinfo {author} {\bibfnamefont {David~I.}\ \bibnamefont {Santiago}}, \bibinfo {author} {\bibfnamefont {Frederic~T.}\ \bibnamefont {Chong}}, \bibinfo {author} {\bibfnamefont {Irfan}\ \bibnamefont {Siddiqi}}, \ and\ \bibinfo {author} {\bibfnamefont {Pranav}\ \bibnamefont {Gokhale}},\ }\bibfield  {title} {\enquote {\bibinfo {title} {Optimized swap networks with equivalent circuit averaging for qaoa},}\ }\href {\doibase 10.1103/PhysRevResearch.4.033028} {\bibfield  {journal} {\bibinfo  {journal} {Phys. Rev. Res.}\ }\textbf {\bibinfo {volume} {4}},\ \bibinfo {pages} {033028} (\bibinfo {year} {2022})}\BibitemShut {NoStop}%
\bibitem [{\citenamefont {Gidney}(2021)}]{Gidney2021stimfaststabilizer}%
  \BibitemOpen
  \bibfield  {author} {\bibinfo {author} {\bibfnamefont {Craig}\ \bibnamefont {Gidney}},\ }\bibfield  {title} {\enquote {\bibinfo {title} {Stim: a fast stabilizer circuit simulator},}\ }\href {\doibase 10.22331/q-2021-07-06-497} {\bibfield  {journal} {\bibinfo  {journal} {{Quantum}}\ }\textbf {\bibinfo {volume} {5}},\ \bibinfo {pages} {497} (\bibinfo {year} {2021})}\BibitemShut {NoStop}%
\bibitem [{Note2()}]{Note2}%
  \BibitemOpen
  \bibinfo {note} {Note that $r_{P,(I,Z)}=r_{P,(Z,I)}$ for the MCM layer we benchmarked, and therefore only 3 analyses are strictly required for each circuit set. We performed both of the equivalent analyses and observed $r_{P,(I,Z)}=r_{P,(Z,I)}$ for all $P$.}\BibitemShut {Stop}%
\bibitem [{\citenamefont {Tripathi}\ \emph {et~al.}(2022)\citenamefont {Tripathi}, \citenamefont {Chen}, \citenamefont {Khezri}, \citenamefont {Yip}, \citenamefont {Levenson-Falk},\ and\ \citenamefont {Lidar}}]{tripathi2022suppression}%
  \BibitemOpen
  \bibfield  {author} {\bibinfo {author} {\bibfnamefont {Vinay}\ \bibnamefont {Tripathi}}, \bibinfo {author} {\bibfnamefont {Huo}\ \bibnamefont {Chen}}, \bibinfo {author} {\bibfnamefont {Mostafa}\ \bibnamefont {Khezri}}, \bibinfo {author} {\bibfnamefont {Ka-Wa}\ \bibnamefont {Yip}}, \bibinfo {author} {\bibfnamefont {E.M.}\ \bibnamefont {Levenson-Falk}}, \ and\ \bibinfo {author} {\bibfnamefont {Daniel~A.}\ \bibnamefont {Lidar}},\ }\bibfield  {title} {\enquote {\bibinfo {title} {Suppression of crosstalk in superconducting qubits using dynamical decoupling},}\ }\href {\doibase 10.1103/PhysRevApplied.18.024068} {\bibfield  {journal} {\bibinfo  {journal} {Phys. Rev. Appl.}\ }\textbf {\bibinfo {volume} {18}},\ \bibinfo {pages} {024068} (\bibinfo {year} {2022})}\BibitemShut {NoStop}%
\bibitem [{\citenamefont {Nielsen}\ \emph {et~al.}(2024)\citenamefont {Nielsen}, \citenamefont {Seritan}, \citenamefont {Ostrove}, \citenamefont {Murray}, \citenamefont {Hines}, \citenamefont {Rudinger}, \citenamefont {Proctor}, \citenamefont {Gamble},\ and\ \citenamefont {Blume-Kohout}}]{pygstiversion0.9.12.3}%
  \BibitemOpen
  \bibfield  {author} {\bibinfo {author} {\bibfnamefont {Erik}\ \bibnamefont {Nielsen}}, \bibinfo {author} {\bibfnamefont {Stefan}\ \bibnamefont {Seritan}}, \bibinfo {author} {\bibfnamefont {Corey}\ \bibnamefont {Ostrove}}, \bibinfo {author} {\bibfnamefont {Riley}\ \bibnamefont {Murray}}, \bibinfo {author} {\bibfnamefont {Jordan}\ \bibnamefont {Hines}}, \bibinfo {author} {\bibfnamefont {Kenny}\ \bibnamefont {Rudinger}}, \bibinfo {author} {\bibfnamefont {Timothy}\ \bibnamefont {Proctor}}, \bibinfo {author} {\bibfnamefont {John}\ \bibnamefont {Gamble}}, \ and\ \bibinfo {author} {\bibfnamefont {Robin}\ \bibnamefont {Blume-Kohout}},\ }\href {http://www.pygsti.info} {\enquote {\bibinfo {title} {Py{GST}i version 0.9.12.3},}\ } (\bibinfo {year} {2024})\BibitemShut {NoStop}%
\bibitem [{\citenamefont {Nielsen}\ \emph {et~al.}(2020)\citenamefont {Nielsen}, \citenamefont {Rudinger}, \citenamefont {Proctor}, \citenamefont {Russo}, \citenamefont {Young},\ and\ \citenamefont {Blume-Kohout}}]{nielsen2020probing}%
  \BibitemOpen
  \bibfield  {author} {\bibinfo {author} {\bibfnamefont {Erik}\ \bibnamefont {Nielsen}}, \bibinfo {author} {\bibfnamefont {Kenneth}\ \bibnamefont {Rudinger}}, \bibinfo {author} {\bibfnamefont {Timothy}\ \bibnamefont {Proctor}}, \bibinfo {author} {\bibfnamefont {Antonio}\ \bibnamefont {Russo}}, \bibinfo {author} {\bibfnamefont {Kevin}\ \bibnamefont {Young}}, \ and\ \bibinfo {author} {\bibfnamefont {Robin}\ \bibnamefont {Blume-Kohout}},\ }\bibfield  {title} {\enquote {\bibinfo {title} {Probing quantum processor performance with py{GST}i},}\ }\href {https://iopscience.iop.org/article/10.1088/2058-9565/ab8aa4} {\bibfield  {journal} {\bibinfo  {journal} {Quantum Sci. Technol.}\ }\textbf {\bibinfo {volume} {5}},\ \bibinfo {pages} {044002} (\bibinfo {year} {2020})}\BibitemShut {NoStop}%
\bibitem [{\citenamefont {Zhang}\ \emph {et~al.}(2024)\citenamefont {Zhang}, \citenamefont {Chen}, \citenamefont {Liu},\ and\ \citenamefont {Jiang}}]{zhang2024generalized}%
  \BibitemOpen
  \bibfield  {author} {\bibinfo {author} {\bibfnamefont {Zhihan}\ \bibnamefont {Zhang}}, \bibinfo {author} {\bibfnamefont {Senrui}\ \bibnamefont {Chen}}, \bibinfo {author} {\bibfnamefont {Yunchao}\ \bibnamefont {Liu}}, \ and\ \bibinfo {author} {\bibfnamefont {Liang}\ \bibnamefont {Jiang}},\ }\href@noop {} {\enquote {\bibinfo {title} {A generalized cycle benchmarking algorithm for characterizing mid-circuit measurements},}\ } (\bibinfo {year} {2024}),\ \Eprint {http://arxiv.org/abs/2406.02669} {arXiv:2406.02669 [quant-ph]} \BibitemShut {NoStop}%
\end{thebibliography}%

\onecolumngrid
\section*{Supplemental Material}
\label{sm}
\section{MCM-CB Theory}
\label{app:theory}
\subsection{MCM-CB with Clifford Gates on Unmeasued Qubits}
In the main text, we assume that the error-free evolution of the unmeasured subystem is $\mathcal{V} = \mathbb{I}_{n-m}$. We obtain the general case, i.e., $\mathcal{V}$ is any $(n-m)$-qubit Clifford operation, by decomposing $\overline{\mathcal{M}} = \overline{\mathcal{M}'}(\mathcal{V}\otimes\mathbb{I}_{m})$, where $\overline{\mathcal{M}'}$ is a USI with ideal evolution $\mathbb{I}_{n-m}$ on the unmeasured subsystem, i.e, 
\begin{equation}
    \overline{\mathcal{M}'}_k =  \sum_{a,b 
    \in \mathbb{Z}_2^m} \mathcal{T}_{a,b}\otimes \mathcal{X}(b) \superketbra{k}{k}\mathcal{X}(a).
\end{equation}
The PTM elements of $\overline{\mathcal{M}}$ are given by
\begin{align}
    \superbra{P' \otimes Z(c_2)}\overline{\mathcal{M}}_k\superket{P \otimes Z(c_1)} = \superbra{P' \otimes Z(c_2)}\overline{\mathcal{M}'}_k\superket{\mathcal{V}[P] \otimes Z(c_1)}.
\end{align}
Because $\mathcal{V}[P]$ is also a Pauli operator, we can now apply Eq. 6 of the main text, and the rest of our theory follows. 
\subsection{Estimating the Process Fidelity of USIs}
In this section, we show that all $\tilde{\lambda}_{P,(Z(c),Z(c'))}$ are linearly independent. We show that any two distinct $\tilde{\lambda}_{P,(Z(c),Z(c'))}$ are orthogonal linear combinations of $\{\lambda_{a,b}\}_{a,b \in \mathbb{Z}_2^m}$ by (1) expressing each $\tilde{\lambda}_{P,(Z(c),Z(c'))}$ as a $2^m$-dimensional vector $\vec{v}_{P,(Z(c),Z(c'))}$ in the basis $\{\lambda_{a,b}\}_{a,b \in \mathbb{Z}_2^m}$, and (2) taking a dot product of the resulting vectors for two parameters $\tilde{\lambda}_{P,(Z(c),Z(c')}$ and $\tilde{\lambda}_{P,(Z(k),Z(k'))}$:
\begin{align}
    \vec{v}_{P,(Z(c),Z(c'))} \cdot \vec{v}_{P,(Z(k), Z(k'))} & = 
    \sum_{a,b \mathbb{Z}_2^m} (-1)^{a \cdot c+ b \cdot c'}(-1)^{a \cdot k + b \cdot k'} \\
    & = \sum_{a,b \mathbb{Z}_2^m} (-1)^{a \cdot (c \oplus k)}(-1)^{b \cdot (c' \oplus k')} \\
    & = \delta_{c, k}\delta_{c,' k'},
\end{align}
where $\delta$ denotes the Kronecker delta. To obtain an expression for $\lambda_{0,0,P}$, we sum over $\tilde{\lambda}_{P,(Z(c),Z(c'))}$ for all $d,c \in \mathbb{Z}_2^m$:
\begin{align} \sum_{c, c' \in \mathbb{Z}_2^m}\tilde{\lambda}_{P,(Z(c),Z(c'))} & = \sum_{c, c' \in \mathbb{Z}_2^m}\sum_{a, b  \in \mathbb{Z}_2^m} \lambda_{a,b,P} (-1)^{a \cdot c + b \cdot c'} \\
& = \sum_{a, b  \in \mathbb{Z}_2^m} \lambda_{a,b,P} \sum_{c,c' \in \mathbb{Z}_2^m}(-1)^{a \cdot c + b \cdot c'}  \\
& = \sum_{a,b \in \mathbb{Z}_2^m} \lambda_{a,a,P} \delta_a\delta_b \\
& = \lambda_{0,0, P}.
\end{align}
Eq. 13 of the main text follows by summing over all $P \in \pauli_{n-m}$.

\subsection{Process Fidelity and Cycle Benchmarking}
\label{app:process_fidelity}
Here, we show that the infinite-sampling limit MCM-CB result $F_{\textrm{MCM-CB}}$. MCM-CB is a lower bound on $F(\overline{\mathcal{M}})$. In the limit of sampling all of the possible MCM-CB subexperiments, the MCM-CB fidelity estimate consists of averaging over all MCM-CB decay constants for a given cycle. The result of this averaging (using the true values of the decay constants), is 
\begin{align}
    F_{\textrm{MCM-CB}} &  = \frac{1}{4^n} \sum_{P \in \mathbf{P}_{n-m}} \sum_{Z(c_1) \in \mathbf{Z}_m} \sum_{Q \in \mathbf{Z}_m}\sqrt[\ell]{\prod_{k=1}^{\ell} \tilde{\lambda}_{\mathcal{V}^k[P], (Z(c_2)^{k-1}Z(c_1), Z(c_2)^{k}Z(c_1)})} \\
    &  \leq \frac{1}{4^n} \sum_{P \in \mathbf{P}_{n-m}} \sum_{Z(c_1) \in \mathbf{Z}_m} \sum_{Q \in \mathbf{Z}_m}\frac{1}{\ell}\sum_{k=1}^{\ell} \tilde{\lambda}_{\mathcal{V}^k[P], (Z(c_2)^{k-1}Z(c_1), Z(c_2)^{k}Z(c_1)}) \\
    &  \leq \frac{1}{4^n} \sum_{P \in \mathbf{P}_{n-m}} \sum_{Z(c_1), Z(c_2) \in \mathbf{Z}_m}\frac{1}{\ell}\sum_{k=1}^{\ell/2} \left(\tilde{\lambda}_{\mathcal{V}^{2k-1}[P], (Z(c_1), Z(c_2))} + \tilde{\lambda}_{\mathcal{V}^{2k}[P], (Z(c_2), Z(c_1))}\right) \\
    &  \leq \frac{1}{4^n} \sum_{P \in \mathbf{P}_{n-m}} \frac{1}{\ell}\sum_{k=1}^{\ell/2} \left(\lambda_{0,0,\mathcal{V}^{2k-1}[P], (Z(c_1), Z(c_2))} + \lambda_{0,0,\mathcal{V}^{2k}[P], (Z(c_2), Z(c_1))}\right) \\
    & \leq \frac{1}{4^n} \sum_{P \in \mathbf{P}_{n-m}} \lambda_{0,0,P}.
\end{align}
This result implies that $F_{\textrm{MCM-CB}} \leq F(\overline{\mathcal{M}})$ for any USI $\overline{\mathcal{M}}$.

\section{MCM-CB Simulations}

\subsection{Simulation Error Models}
\label{app:error models}
Here, we describe the error models used for the simulations of MCM-CB in Fig. 2a. We simulated our method with randomly sampled USIs $\overline{\mathcal{M}} = \mathcal{E}_{\textrm{post}}(\mathcal{T} \otimes \mathbb{I}_{n-m})\mathcal{L} \mathcal{E}_{\textrm{pre}}$. We sampled these USIs so that the infidelity of $\mathcal{T}$ is $p$ and the infidelity of $\mathcal{E}_{\textrm{pre}}$ and $\mathcal{E}_{\textrm{post}}$ are $\nicefrac{p}{2}$, and we sampled $120$ error models with evenly-spaced $p \in [0.0001, 0.0601)$. We modeled initial state preparation and final measurement error as independent bit flip errors on each qubit, which we sampled such that the average state preparation (measurement) error rate per qubit is $0.005$ ($0.01$). 

\subsection{Simulations of MCM-CB on up to 4 Qubits}

\begin{figure*}[t]
    \centering
    \includegraphics{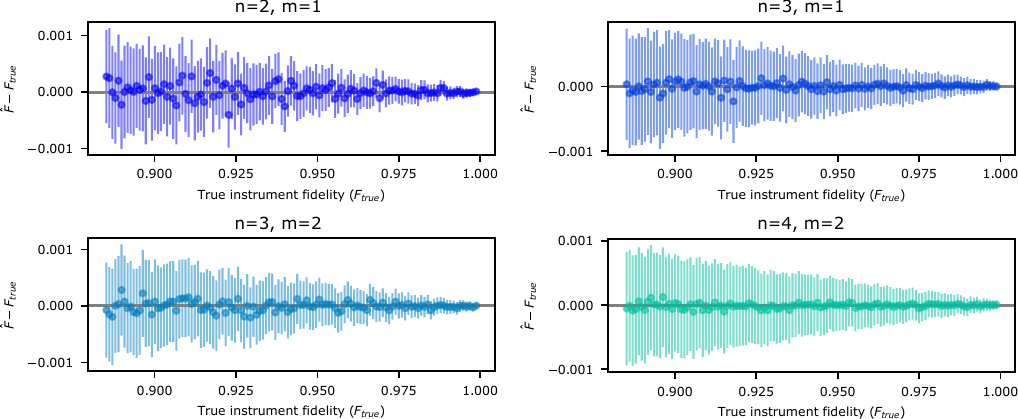}
    \caption{\textbf{Simulating MCM-CB on 2-4 qubits.} The absolute error of MCM-CB fidelity estimates from simulations of MCM-CB with $m=1,2$ measured qubits and $n-m=1,2$ idling qubits, using randomly generated USI error models. The MCM-CB estimate is highly accurate for all error models, and the uncertainty is small.}
    \label{fig:app-few-q}
\end{figure*}

In the few-qubit limit, it is feasible to perform every possible MCM-CB subexperiment, and this is often done in practice for CB of two-qubit gates. We simulated MCM-CB with this exhaustive sampling for USIs with $n-m=1,2$ unmeasured qubits and $m=1,2$ measurements. We simulated our method with randomly sampled USIs using the same sampling parameters as in the $n>4$ qubit simulations in Fig 2a (described in Sec.~\ref{app:error models}). Fig~\ref{fig:app-few-q} shows the fidelity estimates from these simulations. For all error models, the estimated instrument fidelity is within $1\sigma$ of the true fidelity. We observe lower uncertainty in the estimates than observed in our simulations with larger $n$, which we expect due to eliminating uncertainty from sampling MCM-CB subexperiments (note, however, that the uncertainty also depends on the variance of the USI parameters, which is not fixed across our simulations).

\subsection{Simulations of MCM-CB with Additional Error Models}
\label{app:depol}

\begin{figure*}[t]
    \centering
    \includegraphics{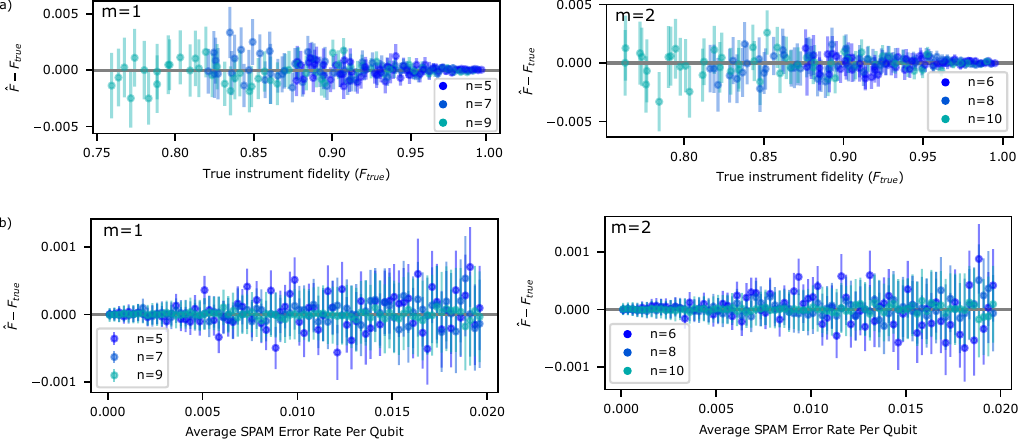}
    \caption{\textbf{Simulating MCM-CB with additional error models.} (a) The absolute error in fidelity estimates from simulated MCM-CB with models consisting of (1) local depolarizing error on each unmeasured qubit, which is independent of error on the measured qubits, and (2) crosstalk errors that include the measured qubits and unmeasured qubits. The MCM-CB estimate is accurate for all error models, with increasing estimate uncertainty with decreasing fidelity. (b) The absolute error in fidelity estimates from simulated MCM-CB with an MCM layer with fidelity $0.94$ and varied rates of initial state preparation and final measurement error. MCM-CB remains accurate for models with average bit flip error rates of up to $2\%$ per qubit.}
    \label{fig:app-depol-spam}
\end{figure*}

The accuracy of the MCM-CB fidelity estimate depends on the properties of the USI error model.  This is because (1) MCM-CB only samples a subset of MCM-CB subexperiments, hence learning only a subset of the USI parameters, and (2) MCM-CB uses a geometric mean to approximate the arithmetic mean of products of USI parameters. To explore this effect, we study the performance of MCM-CB with a different class of error models than in the simulations presented in the main text. We simulated MCM-CB with $n-m=4,6,8$ unmeasured qubits and $m=1,2$ measurements in which we performed $K=100$ MCM-CB subexperiments. We sampled USIs $\overline{\mathcal{M}} = \mathcal{E}_{\textrm{post}}(\mathcal{T} \otimes \mathbb{I}_{n-m})\mathcal{L} \mathcal{E}_{\textrm{pre}}$ where $\mathcal{E}_{\textrm{pre}}$ and $\mathcal{E}_{\textrm{post}}$ each have $100$ random nonzero Pauli error rates, chosen so that the infidelity of $\mathcal{E}_{\textrm{pre}}$ is $5p$ and the infidelity of $\mathcal{E}_{\textrm{post}}$ is $p$. $\mathcal{T}$ consists of single-qubit depolarizing noise on each unmeasured qubit, with the infidelity of each qubit sampled from a normal distribution with mean $p$ and standard deviation $0.2p$. We sample USIs for $120$ uniformly-spaced values $p \in [0.0001, 0.0601)$. We model initial state preparation and final measurement error in the same way as in the simulations in the previous section. Fig~\ref{fig:app-depol-spam} shows the fidelity estimates from these simulations. The MCM-CB estimate is within $2.5\sigma$ of the true fidelity for all error models.  

The simulations presented so far all use a fixed average rate of state preparation and measurement (SPAM) error. To provide further evidence that our method is robust to SPAM error, we also simulated our method with varied-strength SPAM error. We show the results of these simulations in Fig.~\ref{fig:app-depol-spam}(b). We ran MCM-CB with independent bit flip error on each qubit immediately after state preparation and immediately prior to final measurement, and we varied the average per-qubit error rate $p$ from $0$ to $0.02$. To generate the bit flip error rates, for both state preparation and measurement error, we sample uniform random error probabilities and normalize them so that their total is $\nicefrac{pn}{2}$. We observe no systematic effect of the magnitude of the SPAM error on the accuracy of MCM-CB. However, higher SPAM error leads to increased uncertainty in the estimates. 

\subsection{Investigating MCM-CB Subexperiment Sampling in Simulation}
In Fig. 2 of the main text, we simulated MCM-CB with three error models. These models consist of a USI $\overline{\mathcal{M}} = \mathcal{E}_{\textrm{post}}(\mathcal{T} \otimes \mathbb{I}_{n-m})\mathcal{L} \mathcal{E}_{\textrm{pre}}$, and have the following forms:
\begin{enumerate}
    \item (``Sparse'') $\mathcal{T}$, $\mathcal{E}_{\textrm{pre}}$, and $\mathcal{E}_{\textrm{post}}$ have each have $2^{n}$ nonzero Pauli errors, sampled in the same manner as the simulations presented in the main text, such that $\mathcal{T}$ has infidelity $0.04$ and each $\mathcal{E}_{\textrm{pre}}$, and $\mathcal{E}_{\textrm{post}}$ have infidelity $0.01$.
    \item (``Depolarizing+MCM error with crosstalk'') $\mathcal{T}$ consists of single-qubit local depolarizing error, with each qubit's error rate chosen from a normal distribution with mean $0.01$ and standard deviation $0.002$. Each $\mathcal{E}_{\textrm{pre}}$ and $\mathcal{E}_{\textrm{post}}$ have of $2^{n}$ randomly sampled nonzero Pauli error rates. $\mathcal{E}_{\textrm{pre}}$ and $\mathcal{E}_{\textrm{post}}$ are sampled to have the same fidelity ($0.01$).
    \item (``Depolarizing+pre-MCM error with crosstalk'') $\mathcal{T}$ consists of single-qubit local depolarizing error sampled from a normal distribution with mean $0.005$ and standard deviation $0.0001$, $\mathcal{E}_{\textrm{pre}}$ consists of $2^n$ randomly-sampled Pauli errors, and $\mathcal{E}_{\textrm{post}} = \mathbb{I}_n$ with total infidelity $0.02$.
\end{enumerate}

\section{Layer Set Cycle Benchmarking}
\label{app:cz}
\subsection{Learnability of Layer Set Parameters}

In this section, we explain how to learn all non-gauge parameters of a layer set using layer set cycle benchmarking. The learnable parameters of a layer set correspond to the parameters in the \emph{cycle space} of the layer set's pattern transfer graph (PTG) \cite{chen_learnability_2023}. Each edge $e : v \rightarrow v'$ of the PTG is a parameter of a layer in the layer set, and we denote this layer by $\ell(e)$. Each edge also has two associated Pauli operators, $\mathrm{PRE}(e)$ and $\mathrm{POST}(e)$, where the support of $\mathrm{PRE}(e)$ is encoded by $v$ and the support of $\mathrm{POST}(e)$ is encoded by $v'$. Traversing an edge from vertex $v$ to vertex $v'$ via edge $e$ corresponds to applying $\ell(v)$ and an associated processing of any MCM results, encoded by bit string $t_{e} \in \mathbb{Z}_2^m$. The next effect of these operations is to transform $\mathrm{PRE}(e)$ into $\mathrm{POST}(e)$. 

Given a cycle in the PTG specified by a sequence of edges $\{e_1, \cdots e_k\}$ (where we have arbitrarily chosen a starting edge), the MCM-CB circuits to learn the parameter $\prod_{i=1}^ke_i$ have the form $\layer{S}_0\layer{G}^d\layer{S}_f$, where $d \geq 0$ and
\begin{enumerate}
    \item $\layer{S}_0$ is a layer of single-qubit gates preparing an eigenstate of $\mathrm{PRE}(e_1)$.
    \item $G = \layer{S}_k\layer{L}_k \cdots \layer{S}_1\layer{L}_1$, where $\layer{L}_i=\ell(e_i)$, and $\layer{S}_i$ is a layer of single qubit gates that transforms $\mathrm{POST}(e_i)$ into $\mathrm{PRE}(e_{i+1})$ (and we define $e_{k+1}\equiv e_1$. Such an $S_i$ must exist, because the support of $\mathrm{POST}(e_i)$ is the same as that of $\mathrm{PRE}(e_{i+1})$.
    \item $\layer{S}_{f}$ is a layer of gates that transforms $\mathrm{POST}(e_{k})$ into a Z-type Pauli. 
\end{enumerate}
For each circuit, we measure 
\begin{equation}
f(\layer{C}) = (-1)^{t_f \cdot b_f+t_0}\prod_{i=1}^{d}\prod_{j=1}^{k} (-1)^{b_{j,i} \cdot t_{e_j, i}}, \label{eq:lscb_data_analysis}
\end{equation}
where $b_{j,i}$ is the MCM outcomes of layer $j$ in the $i$th repetition of $G$, $t_{e_j, i}$ is the MCM processing string encoded in $e_j$, and $b_f, t_f,$ and $t_0$ are as defined in the main text. 

This construction shows that MCM-CB is capable of learning every layer set parameter in the cycle space of the PTG. Furthermore, these parameters are \emph{all} of the non-gauge parameters of the layer set, which can be shown by constructing a gauge transformation between any two elements of the cut space (i.e., the complement of the cycle space) of the PTG \cite{zhang2024generalized, chen_learnability_2023}.

\subsection{Layer Set Cycle Benchmarking with CZ and an MCM}
In this section, we expand on the example of performing MCM-CB with a 2-qubit layer set consisting of two layers: a controlled $Z$ (CZ) gate (between two qubits \Q{0} and \Q{1}) and a single-qubit MCM (which we will assume is on \Q{0}, but the case of an MCM on \Q{1} is analogous). To clarify the position of the MCM, we will use the notation $\tilde{\lambda}_{(Z(c_1),Z(c_2)),P}$ to denote the PTM elements of the USI of the MCM layer. We model the CZ gate's error as a post-gate stochastic Pauli channel with eigenvalues $\lambda_Q^{CZ}$ and Pauli error rates $p_Q$  for $Q \in \mathbf{P}_2$. The pattern transfer graph for this layer set is shown in Fig.~\ref{fig:ptm}. The cycle space of this graph fully describes the learnable parameters of this layer set for MCM-CB with arbitrary single-qubit gates between layers \cite{chen_learnability_2023}, which are as follows.
\begin{itemize}
\item Learnable with MCM-only CB: $\tilde{\lambda}_{(Z,I),I}\tilde{\lambda}_{(I,Z),I}$, $\tilde{\lambda}_{(I,I),P}$, and $\tilde{\lambda}_{(Z,Z),P}$, and $\tilde{\lambda}_{(Z,I),P'}\tilde{\lambda}_{(I,Z),P}$ for $P,P;' =  X, Y, Z$
\item Learnable with CZ-only CB: $\lambda^{CZ}_{II}$, $\lambda^{CZ}_{IZ}$, $\lambda^{CZ}_{ZI}$, $\lambda^{CZ}_{XY}$, $\lambda^{CZ}_{YY}$, $\lambda^{CZ}_{YX}$, $\lambda^{CZ}_{XX}$, $\lambda^{CZ}_{ZZ}$, $\lambda^{CZ}_{IX}\lambda^{CZ}_{ZX}$, $\lambda^{CZ}_{IY}\lambda^{CZ}_{ZX}$, $\lambda^{CZ}_{IX}\lambda^{CZ}_{ZY}$,
$\lambda^{CZ}_{IY}\lambda^{CZ}_{ZY}$,
$\lambda^{CZ}_{XI}\lambda^{CZ}_{XZ}$, $\lambda^{CZ}_{YI}\lambda^{CZ}_{XZ}$, $\lambda^{CZ}_{XI}\lambda^{CZ}_{YZ}$, and
$\lambda^{CZ}_{YI}\lambda^{CZ}_{YZ}$,
\item Learnable with CZ+MCM CB: $\lambda^{CZ}_{ZY}\tilde{\lambda}_{P,(Z,I)}$, $\lambda^{CZ}_{IY}\tilde{\lambda}_{P,(I,Z)}$, $\lambda^{CZ}_{ZX}\tilde{\lambda}_{P,(Z,I)}$, $\lambda^{CZ}_{IX}\tilde{\lambda}_{P,(I,Z)}$ for $P = X, Y, Z$
\end{itemize}

Parameters learned by CB of a CZ and MCM sequence are \emph{relational errors} that cannot be separated into CZ and MCM-only errors. As such, they can be difficult to interpret. Furthermore, we cannot perform a Walsh-Hadamard transform the relation errors into sums of Pauli error rates. As a step towards interpreting these error rates, we perform a first order expansion of these parameters in the Pauli error rates. First, we express $\tilde{\lambda}_{(Z,I),P}$, $\tilde{\lambda}_{(I,Z),P}$ exactly in terms of Pauli error rates:
\begin{align}
    \tilde{\lambda}_{(Z,I),P} & = \lambda_{0,0,P}-\lambda_{1,0,P}+\lambda_{0,1,P}-\lambda_{1,1,P} \\
    & = p_{0,0} - p_{1,0} + p_{0,1} - p_{1,1} -2\left(\sum_{P': [P,P'] \neq 0} p_{0,0,P}-p_{1,0,P}+p_{0,1,P}-p_{1,1,P}\right) \\ 
    & = 1-2(p_{0,1}+p_{1,1})-2\left(\sum_{P': [P,P'] \neq 0} p_{0,0,P}-p_{1,0,P}+p_{0,1,P}-p_{1,1,P}\right),
\end{align}
where $p_{a,b}$ denotes the probability of pre-measurement MCM error $\mathcal{X}(a)$ and post-measurement error $\mathcal{X}(b)$, and $p_{a,b,P}$ denotes the probability of pre-measurement error $\mathcal{X}(a)$, post-measurement error $\mathcal{X}(b)$, and error $P$ on the unmeasured qubits. Analogously, 
\begin{align}
    \tilde{\lambda}_{(Z,I),P} = 1-2(p_{1,0}+p_{1,1})-2\left(\sum_{P': [P,P'] \neq 0} p_{0,0,P}+p_{1,0,P}-p_{0,1,P}-p_{1,1,P}\right),
\end{align}
We now compute the CZ-MCM relational parameters and drop second order terms, 
\begin{align}
    \lambda^{CZ}_{ZX}\tilde{\lambda}_{(Z,I),P} & \approx 1 - 2(p_{XI}+p_{XX}+p_{YI}+p_{YX}+p_{ZY}+p_{ZZ}+p_{IY}+p_{IZ}) -2(p_{1,0}+p_{1,1})-2\left(\sum_{P': [P,P'] \neq 0} p_{0,0,P'} - p_{1,0,P'} + p_{0,1,P'}-p_{1,1,P'}\right) \\
    & = 1- 2(p_{XI}+p_{XX}+p_{YI}+p_{YX}+p_{ZY}+p_{ZZ}+p_{IY}+p_{IZ}) -2(p_{1,0,I}+p_{1,1,I}+p_{1,0,P}+p_{1,1,P})-2\left(\sum_{P': [P,P'] \neq 0} p_{0,0,P'} + p_{0,1,P'}\right) \\
    \lambda^{CZ}_{ZY}\tilde{\lambda}_{(Z,I),P} & \approx 1 - 2(p_{XI}+p_{XY}+p_{YI}+p_{YY}+p_{ZX}+p_{ZZ}+p_{IX}+p_{IZ}) -2(p_{1,0}+p_{1,1})-2\left(\sum_{P': [P,P'] \neq 0} p_{0,0,P'} - p_{1,0,P'} + p_{0,1,P'}-p_{1,1,P'}\right) \\
    & = 1 - 2(p_{XI}+p_{XY}+p_{YI}+p_{YY}+p_{ZX}+p_{ZZ}+p_{IX}+p_{IZ}) -2(p_{1,0, I}+p_{1,1,I}+p_{1,0, P}+p_{1,1,P})-2\left(\sum_{P': [P,P'] \neq 0} p_{0,0,P'}  + p_{0,1,P'}\right) \\
    \lambda^{CZ}_{IX}\tilde{\lambda}_{(I,Z),P} & \approx 1 - 2(p_{IY}+p_{IZ}+p_{XY}+p_{XZ}+p_{YY}+p_{YZ}+p_{ZY}+p_{ZZ}) -2(p_{0,1}+p_{1,1})-2\left(\sum_{P': [P,P'] \neq 0} p_{0,0,P'} + p_{1,0,P'} - p_{0,1,P'}-p_{1,1,P'}\right) \\
    & = 1 - 2(p_{IY}+p_{IZ}+p_{XY}+p_{XZ}+p_{YY}+p_{YZ}+p_{ZY}+p_{ZZ}) -2(p_{0,1,I}+p_{1,1,I}+p_{0,1,P}+p_{1,1,P})-2\left(\sum_{P': [P,P'] \neq 0} p_{0,0,P'} + p_{1,0,P'}\right) \\
    \lambda^{CZ}_{IY}\tilde{\lambda}_{(I,Z),P} & \approx 1 - 2(p_{IX}+p_{IZ}+p_{XX}+p_{XZ}+p_{YX}+p_{YZ}+p_{ZX}+p_{ZZ}) -2(p_{0,1}+p_{1,1})-2\left(\sum_{P': [P,P'] \neq 0} p_{0,0,P'} + p_{1,0,P'} - p_{0,1,P'}-p_{1,1,P'}\right) \\
    & = 1 - 2(p_{IX}+p_{IZ}+p_{XX}+p_{XZ}+p_{YX}+p_{YZ}+p_{ZX}+p_{ZZ}) -2(p_{0,1,I}+p_{1,1,I}+p_{0,1,P}+p_{1,1,P})-2\left(\sum_{P': [P,P'] \neq 0} p_{0,0,P'} + p_{1,0,P'}\right)
\end{align}
We see that by performing CB with the CZ+MCM germ, we do not learn $p_{0,1,P'}$ $p_{1,0,P'}$ in isolation, but we learn information about $p_{0,1,P'}-p_{1,0,P'}$ in combination with many CZ error rates. 

It is also useful to expand the parameters learned by MCM-CB of the MCM layer to first order. This offers an alternate approach to estimating Pauli error probabilities using MCM-CB data, instead of the approximation $\tilde{\lambda}_{(I,Z), P} \approx \tilde{\lambda}_{(Z,I), P} \approx \hat{p}_{(I,Z),P}$ used in the main text. To first order, the parameters we learn in product are 
\begin{align}
    \tilde{\lambda}_{(Z,I),P}\tilde{\lambda}_{(I,Z),P} & = 1 - 2(p_{0,1} + p_{1,0} + 2p_{1,1}) - 4\left(\sum_{P': [P,P'] \neq 0} p_{0,0,P'}-p_{1,1,P'}\right) \\
    \tilde{\lambda}_{(Z,I),I}\tilde{\lambda}_{(I,Z),I} & = 1 - 2\left(p_{0,1} + p_{1,0} + 2p_{1,1}\right).
\end{align}
The second equation allows us to learn $p_{1,1}$ and $p_{0,0}$, because we can learn $p_{0,1}+p_{1,0}$ from $\tilde{\lambda}_{(I,I),I}$ and $\tilde{\lambda}_{(Z,Z),I}$. Furthermore, to first order, 
\begin{align}
    \tilde{\lambda}_{(Z,I),I}\tilde{\lambda}_{(I,Z),I}-\tilde{\lambda}_{(Z,I),P}\tilde{\lambda}_{(I,Z),P} = 4\left(\sum_{P': [P,P'] \neq 0} p_{0,0,P'}-p_{1,1,P'}\right)
\end{align} 
Using the above equation, we can estimate $p_{0,0,P}$, $p_{1,1,P}$ and $p_{0,1,P}+p_{1,0,P}$ for $P=X,Y,Z$ using $\tilde{\lambda}_{(Z,I),I}\tilde{\lambda}_{(I,Z),I}$ and all $\tilde{\lambda}_{(Z,I),P}\tilde{\lambda}_{(I,Z),P}$, $\tilde{\lambda}_{(Z,Z),P}$ and $\tilde{\lambda}_{(I,I),P}$.

\begin{figure}
\label{fig:ptm}
  \begin{tikzpicture}

    \node[vertex] (1) {$00$};
    \node[vertex] (2) [right = 3cm  of 1]  {$01$};
    \node[vertex] (3) [below = 3cm  of 1] {$10$};
    \node[vertex] (4) [right = 3cm  of 3] {$11$};

    \draw[edge] (2) to [bend left=40]  node [midway, right] {$\tilde{\lambda}_{(I,Z),P}$} (4);
    \draw[edge] (4) to [bend left=40]  node [midway, left] {$\tilde{\lambda}_{(Z, I),P}$} (2);
    \draw[edge] (1) to [bend left] node [midway, right] {$\tilde{\lambda}_{(I,Z),P}$} (3);
    \draw[edge] (3) to [bend left]  node [midway, left] {$\tilde{\lambda}_{(Z, I),P}$} (1);
    \draw[edge] (4) to [bend left=5] node [midway, left, rotate=90, anchor=south] {$\lambda_{YZ}^{CZ},\lambda_{XZ}^{CZ}$} (2) ;
    \draw[edge] (3) to [bend left=5] node  [midway, above] {$\lambda_{YI}^{CZ},\lambda_{XI}^{CZ}$} (4) ;
    \draw[edge] (2) to [bend left=5] node [midway, right, rotate=270, anchor=south] {$\lambda_{IY}^{CZ},\lambda_{IX}^{CZ}$} (4) ;
    \draw[edge] (4) to [bend left=5] node [midway, below] {$\lambda_{ZY}^{CZ},\lambda_{ZX}^{CZ}$} (3) ;
    \draw[edge] (1) to [loop left] node [midway, left] {$\tilde{\lambda}_{(I, I),I}$} (1);
    \draw[edge] (4) to [loop right] node [midway, right]  {$\tilde{\lambda}_{(Z,Z),P}$} (4);
    \draw[edge] (3) to [loop left] node [midway, left] {$\tilde{\lambda}_{(Z, Z),I}$} (3);
    \draw[edge] (2) to [loop right] node [midway, right] {$\tilde{\lambda}_{(I, I),P}$} (2);
    \draw[edge] (1) to [loop above] node [midway, left] {$\lambda_{II}^{CZ}$} (1);
    \draw[edge] (4) to [loop below] node [midway, right] {$\lambda_{XY}^{CZ}$,$\lambda_{YY}^{CZ}$,$\lambda_{YX}^{CZ}$,$\lambda_{XX}^{CZ}$,$\lambda_{ZZ}^{CZ}$} (4);
    \draw[edge] (3) to [loop below] node [midway, right] {$\lambda_{ZI}^{CZ}$} (3);
    \draw[edge] (2) to [loop above] node [midway, right] {$\lambda_{IZ}^{CZ}$} (2);
\end{tikzpicture}
\caption{\textbf{MCM-CB of MCM+CZ.} Pattern transfer graph for a gate set consisting of single-qubit MCMs and a CZ gate. We label the edges with bit strings denoting the support of Pauli operators (e.g., $01$ corresponds to the Pauli operators $IX, IY,$ and $IZ$). We use $P$ to denote any non-identity single-qubit Pauli operator.}
\end{figure}

\section{Details of IBM Q Demonstrations}
\label{app:ibmq}

\begin{figure*}[h]
    \centering    \includegraphics{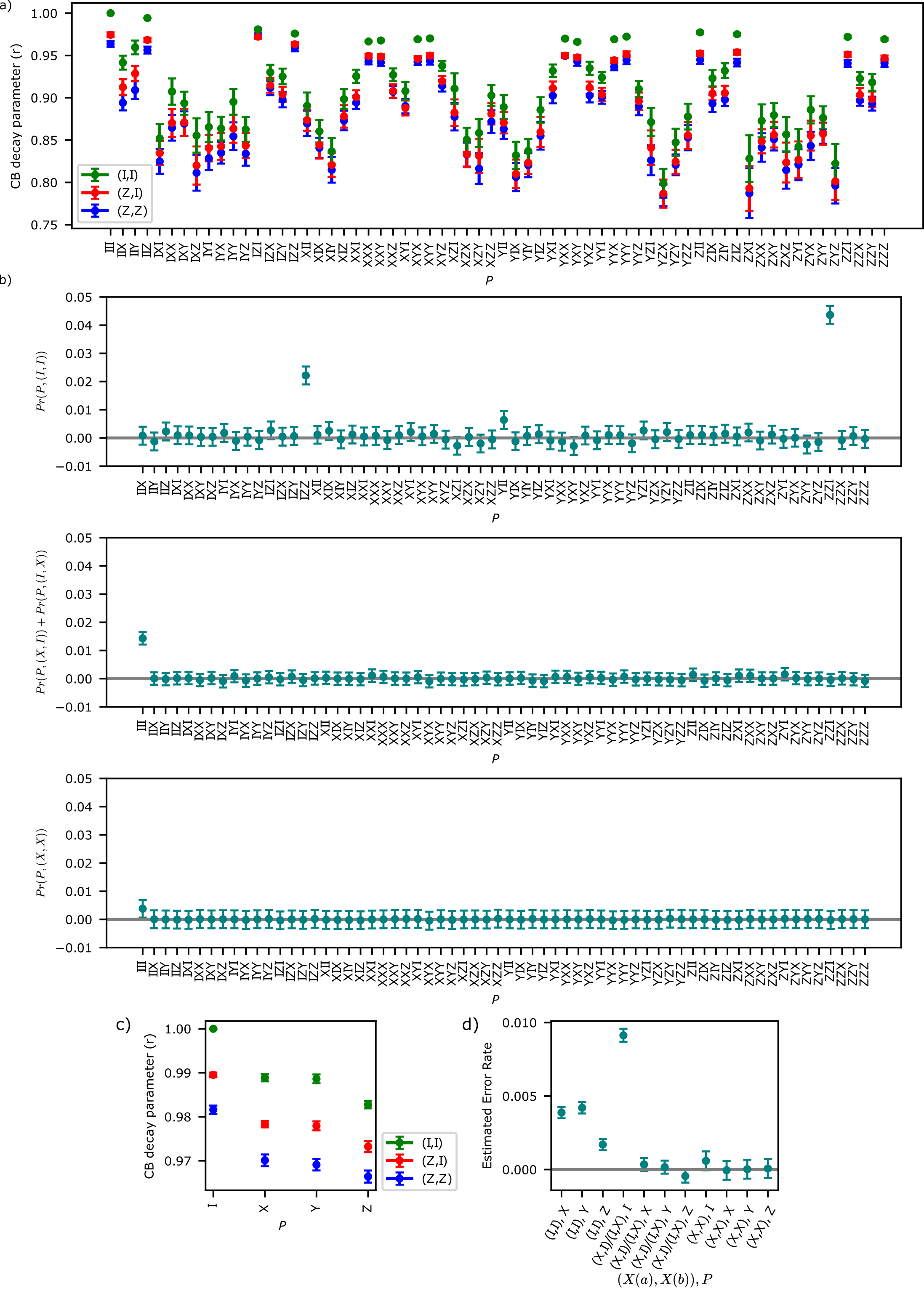}
    \caption{\textbf{MCM-CB on IBM Q.} (a) All extracted MCM-CB parameters $\hat{r}_{P,(Z(c_1),Z(c_2))}$ from our demonstration of MCM-CB of a 4-qubit layer with one MCM on \texttt{ibm\_osaka}. (b) The estimated Pauli error rates, which we obtain by performing a Walsh-Hadamard transformation on subsets of the MCM-CB decay parameters with fixed $(Z(c_1),Z(c_2))$ and using the approximation $r_{P,(Z(c_1),Z(c_2))} \approx \tilde{\lambda}_{P,(Z(c_1),Z(c_2))} \approx \tilde{\lambda}_{P,(Z(c_2),Z(c_1))}$ for $c_1 \neq c_2$.  (c) All extracted parameters from 2-qubit CB MCM-CB of a single-qubit MCM on \texttt{ibm\_torino}. The estimated fidelity of the MCM layer is $0.980(1)$. (d) Estimated Pauli error rates for the 2-qubit MCM layer.}
    \label{fig:app_ibmq}
\end{figure*}

In this section, we provide additional details of our demonstrations of MCM-CB on IBM Q processors. Figure~\ref{fig:app_ibmq}(a) shows the MCM-CB decay parameters $\hat{r}_{P,(Z(c_1),Z(c_2))}$ from our 4-qubit MCM-CB demonstration on \texttt{ibm\_osaka}. We omit the decay parameters $\hat{r}_{P,(I,Z)}$ because $r_{P,(I,Z)}=r_{P,(Z,I)}$, and our estimates of $r_{P,(I,Z)}$ and $r_{P,(I,Z)}$ were approximately equal. We estimated the eigenvalues of the $\mathcal{T}_{a,b}$ using
\begin{align}
    \hat{\lambda}_{0,0,P} & = \frac{1}{4}(\hat{p}_{P, (I, I)} + \hat{p}_{P, (Z, Z)} + 2\hat{p}_{P, (Z, I)}) \\    \hat{\lambda}_{0,1,P}+\hat{\lambda}_{1,0,P} & = \frac{1}{2}(\hat{p}_{P, (I, I)} - \hat{p}_{P, (Z, Z)}) \\
    \hat{\lambda}_{1,1,P} & =\frac{1}{4}( \hat{p}_{P, (I, I)} + \hat{p}_{P, (Z, Z)} - 2\hat{p}_{P, (Z, I)}).
\end{align}
Note that our estimates for $\lambda_{0,0,P}$, $\lambda_{1,1,P}$ use the approximation $\tilde{\lambda}_{P,(I,Z)}\approx \tilde{\lambda}_{P,(Z,I)} \approx r_{P,(Z,I)}$ for all $P$.
We then perform a Walsh-Hardamard transform on each of the sets of values $\{\hat{\lambda}_{0,0,P}\}$, $\{\hat{\lambda}_{1,1,P}\}$, and $\{\hat{\lambda}_{0,1,P}+\hat{\lambda}_{1,0,P}\}$ to estimate the Pauli error rates of the MCM layer \cite{harper2020efficient}. Figure~\ref{fig:app_ibmq}(b) shows all estimated Pauli error rates. Table~\ref{fig:ibm_osaka_calibration} shows the calibration data for \texttt{ibm\_osaka}  from the time we ran our MCM-CB circuits.

In addition to our 4-qubit demonstration, we ran 2-qubit MCM-CB of a layer consisting of an MCM (on \Q{13}) and an idling qubit (\Q{14}) of \texttt{ibm\_torino}. Figure~\ref{fig:app_ibmq}(c) shows all MCM-CB decay parameters from this demonstration, and Fig.~\ref{fig:app_ibmq}(d) shows all estimated Pauli error rates for the MCM layer, using the same approach as in our 4-qubit MCM-CB demonstration to perform the estimation. For this layer, single MCM bit flip errors are the dominant source of error [0.009(1)], followed by $Y$, $X$, and $Z$ errors on the unmeasured qubit. The calibration data from the time of execution is shown in Table~\ref{app:ibm_torino_calibration}.

\begin{table*}[h]
\begin{tabular}{|l|l|l|l|l|l|l|l|l|}
\cline{1-9}
qubit & $T_1$ (us) & $T_2$ (us) & frequency (GHz)& anharmonicity  (GHz) & readout error & Pr(prep 1, measure 0)& Pr(prep 0, measure 1)& readout length  (ns) \\\cline{1-9}
\Q40 & 214.84 & 7.93 & 4.89 & -0.31 & 0.006 & 0.006 & 0.007 & 1400.00 \\\cline{1-9}
\Q41 & 188.37 & 101.57 & 4.99 & -0.31 & 0.040 & 0.044 & 0.036 & 1400.00 \\\cline{1-9}
\Q42 & 226.91 & 5.93 & 4.79 & -0.31 & 0.010 & 0.012 & 0.007 & 1400.00 \\\cline{1-9}
\Q43 & 409.52 & 135.55 & 4.93 & -0.31 & 0.018 & 0.006 & 0.030 & 1400.00 \\\cline{1-9}
\end{tabular}
\caption{\textbf{IBM Q Calibration data for \texttt{ibm\_osaka}.} Calibration data from the time of our 4-qubit MCM-CB  demonstration.}
\label{fig:ibm_osaka_calibration}
\end{table*}

\begin{table*}[h]
\begin{tabular}{|l|l|l|l|l|l|l|l|l|}
\cline{1-9}
qubit & $T_1$ (us) & $T_2$ (us) & frequency (GHz)& anharmonicity  (GHz) & readout error & Pr(prep 1, measure 0)& Pr(prep 0, measure 1)& readout length  (ns) \\\cline{1-9}
\Q13 & 288.34 & 93.44 & 4.73 & 0.00 & 0.008 & 0.012 & 0.003 & 1400.00 \\\cline{1-9}

\Q14 & 129.27 & 29.80 & 4.74 & -0.31 & 0.085 & 0.091 & 0.079 & 1400.00 \\\cline{1-9}
\end{tabular}
\caption{\textbf{IBM Q Calibration data for \texttt{ibm\_torino}.} Calibration data from the time of our 2-qubit MCM-CB  demonstration.}
\label{app:ibm_torino_calibration}
\end{table*}
\end{document}